\RequirePackage{fix-cm}
\documentclass[twocolumn]{svjour3}          
\smartqed  

\usepackage{amsthm}
\usepackage{amsmath}
\usepackage{bm}
\usepackage{natbib}
\usepackage[toc,page]{appendix}
\usepackage[colorlinks,citecolor=blue,urlcolor=blue,filecolor=blue,backref=page]{hyperref}
\usepackage[linesnumbered, algoruled, longend, boxed]{algorithm2e}
\usepackage{multirow}
\usepackage{graphicx}
\usepackage{mathrsfs}
\usepackage{float}
\usepackage{subfigure}
\usepackage{siunitx}
\usepackage{xcolor}

\newcommand{\mmu}{\bm{\mathit{\mu}}}
\newcommand{\mtheta}{\mathbf{\mathit{\theta}}}
\newcommand{\mTheta}{\bm{\theta}}
\newcommand{\mXi}{\bm{\xi}}
\newcommand{\mSigma}{\bm{\Sigma}}

\newcommand{\mPr}{\Pr}
\newcommand{\mL}{\mathcal{L}}

\newcommand{\mpi}{\mathit{\pi}}

\newcommand{\mZ}{\mathcal{Z}}

\newcommand{\mP}{\mathcal{P}}

\newcommand{\ml}{\mathit{l}}

\newcommand{\mxi}{\mathit{\xi}}

\newcommand{\mV}{\mathit{V}}
\newcommand{\mN}{\mathit{N}}

\newcommand{\tmL}{\tilde{\mathcal{L}}}
\newcommand{\tmpi}{\tilde{\mathit{\pi}}}

\newcommand{\mbeta}{\mathit{\beta}}

\newcommand{\calD}{\mathcal{D}}
\newcommand{\calM}{\mathcal{M}}
\newcommand{\calN}{\mathcal{N}}
\newcommand{\calU}{\mathcal{U}}

\newcommand{\revised}{\textcolor{black}}

\makeatletter
\newcommand{\mypm}{\mathbin{\mathpalette\@mypm\relax}}
\newcommand{\@mypm}[2]{\ooalign{%
  \raisebox{.3\height}{$#1+$}\cr
  \smash{\raisebox{-.4\height}{$#1-$}}\cr}}
\makeatother

\graphicspath{{img/},{}}
\journalname{Statistics and Computing}
\begin{document}

\title{
Improving the efficiency and robustness of
nested sampling using posterior repartitioning
}


\author{Xi Chen    \and
        Michael Hobson   \and
                Saptarshi Das  \and
        Paul Gelderblom 
}


\institute{X. Chen, M. Hobson, and S. Das are with Cavendish Laboratory, Department of Physics, University of Cambridge, UK. \email{xc253@cam.ac.uk,  mph@mrao.cam.ac.uk, sd731@mrao.cam.ac.uk}.           
           \and  \\
           P. Gelderblom is with Shell Global Solutions International BV, Netherlands. 
           \email{paul.gelderblom@shell.com.}
}

\date{Received:  / Accepted: }

\maketitle

\begin{abstract}
In real-world Bayesian inference applications, prior assumptions
regarding the parameters of interest may be unrepresentative of their
actual values for a given dataset. In particular, if the likelihood is
concentrated far out in the wings of the assumed prior distribution,
this can lead to extremely inefficient exploration of the resulting
posterior by nested sampling (NS) algorithms, with unnecessarily high
associated computational costs. Simple solutions such as broadening
the prior range in such cases might not be appropriate or possible in
real-world applications, for example when one wishes to assume a
single standardised prior across the analysis of a large number of
datasets for which the true values of the parameters of interest may
vary. This work therefore introduces a posterior repartitioning (PR)
method for NS algorithms, which addresses the problem by
redefining the likelihood and prior while keeping their product fixed,
so that the posterior inferences \revised{and evidence estimates} remain unchanged but the efficiency
of the NS process is significantly increased. Numerical
results show that the PR method provides a simple yet powerful
refinement for NS algorithms to address the issue of
unrepresentative priors.  \keywords{Bayesian modelling \and nested
  sampling \and unrepresentative prior \and posterior repartitioning}
\end{abstract}

\section{Introduction}

Bayesian inference (see e.g. \citealt{mackay2003}) provides a
comprehensive framework for estimating unknown parameter(s) $\mtheta$
of some model with the assistance both of
observed data $\calD$ and prior knowledge of $\mtheta$. One is
interested in obtaining the posterior distribution of $\mtheta$, and
this can be expressed using Bayes' theorem as:
\begin{align}
\mPr(\mtheta | \calD, \calM) = \frac{\mPr(\calD | \mtheta, \calM) \mPr(\mtheta | \calM)}{\mPr(\calD | \calM)},
\end{align}
where $\calM$ represents model (or hypothesis) assumption(s),
$\mPr(\mtheta | \calD, \calM) \equiv \mP(\mtheta)$ is the
\texttt{posterior} probability density, $\mPr(\calD | \mtheta, \calM)
\equiv \mL(\mtheta)$ is the \texttt{likelihood}, and $\mPr(\mtheta |
\calM) \equiv \mpi(\mtheta)$ is the \texttt{prior} of
$\mtheta$. $\mPr(\calD | \calM) \equiv \mZ$ is called the
\texttt{evidence} (or marginal likelihood). We then have a simplified
expression:
\begin{align}
\mP(\mtheta) = \frac{\mL(\mtheta) \mpi(\mtheta)}{\mZ},
\label{Eq:Bayes}
\end{align} 
and 
\begin{align}
\mZ = \int_{\Psi} \mL(\mtheta)  \mpi(\mtheta) d \mtheta, 
\label{Eq:origZ}
\end{align}
\revised{where $\Psi$ represents the prior space of $\mtheta$.} The evidence $\mZ$ is often used for model selection. It is the
average of the likelihood over the prior, considering every possible
choice of $\mtheta$, and thus is not a function
of the parameters $\mtheta$. By ignoring the constant
$\mZ$, the posterior $\mP(\mtheta)$ is proportional to the product of
likelihood $\mL(\mtheta)$ and prior $\mpi(\mtheta)$.

The likelihood $\mL(\mtheta)$ is fully determined by the observation
model (or measurement model / forward model) along with its
corresponding noise assumptions. It is common that the structure of the
observation model is predefined in real-world applications. By
contrast, the prior distribution is often less well defined, and can
be chosen in a number of ways, provided it is consistent with any
physical requirements on the parameters $\mtheta$ (or quantities
derived therefrom).  One role of the prior distribution
$\mpi(\mtheta)$ is to localise the appropriate region of interest in
the parameter space, which assists the inference process. One often
chooses a standard distribution (such as Gaussian or uniform) as the
prior when limited information is available {\em a priori}. In particular, the
prior should be representative of the range of values that the
parameters might take for the dataset(s) under analysis. An interesting
discussion related to prior belief in a broader context can be
found in \cite{gelman2008}.

The approach outlined above works well in most scenarios, but it can
be problematic if an inappropriate prior is chosen. In particular, if
the true values of the parameters $\mtheta$ [or, more meaningfully,
  the location(s) of the peak(s) of the likelihood] lie very far out
in the wings of the prior distribution $\mpi(\mtheta)$, then this can
result in very inefficient exploration of the parameter space by
NS algorithms. In extreme cases, it can even result in a
sampling algorithm failing to converge correctly, usually because of
numerical inaccuracies, and incorrect posterior inferences (a toy
example will be used to illustrate this problem in later sections).

This paper seeks to address the \texttt{unrepresentative prior}
problem.  One obvious solution is simply to augment the prior so that
it covers a wider range of the parameter space.  In some common cases,
however, this might not be applicable. This is particularly true when
one wishes to assume the same prior across a large number of datasets,
for each of which the peak(s) of the likelihood may lie in very
different regions of the parameter space.  Moreover, in practical
implementations, the specialists responsible for defining the prior
knowledge, developing the measurement model, building the software,
performing the data analysis, and testing the solution are often
different people. Thus, there may be a significant overhead in
communicating and understanding the full analysis pipeline before a
new suitable prior could be agreed upon for a given scenario. This is
a common occurrence in the analysis of, for example, production data
in the oil and gas industry. 

We therefore adopt an approach in this paper that circumvents the
above difficulties. In particular, we present a posterior repartitioning
(PR) method for addressing the unrepresentative prior problem in the
context of NS algorithms \citep{skilling2006nested} for
exploring the parameter space. One important way in which nested
sampling differs from other methods is that it makes use of the
likelihood $\mL(\mtheta)$ and prior $\pi(\mtheta)$ {\em separately} in
its exploration of the parameter space\revised{, in that samples are drawn from the prior $\pi(\mtheta)$ such that they satisfy some likelihood constraint $\mL(\mtheta) > L_\ast$.} By contrast, Markov chain Monte Carlo (MCMC) sampling methods or genetic algorithm variants are
typically blind to this separation\footnote{One exception is
  the propagation of multiple MCMC chains, for which it is often
  advantageous to draw the starting point of each chain independently
  from the prior distribution.}, and deal solely in terms of
the product $\mL(\mtheta)\pi(\mtheta)$, which is proportional to the
posterior $\mP(\mtheta)$. This difference provides an opportunity in
the case of NS to `repartition' the product
$\mL(\mtheta)\pi(\mtheta)$ by defining a new effective likelihood
$\tmL(\mtheta)$ and prior $\tmpi(\mtheta)$ (which is typically
`broader' than the original prior), subject to the
condition $\tmL(\mtheta)\tmpi(\mtheta)= \mL(\mtheta)\pi(\mtheta)$, so
that the (unnormalised) posterior remains unchanged. Thus, in
principle, the inferences obtained are unaffected by the use of the PR
method, but, as we will demonstrate, the approach can yield
significant improvements in sampling efficiency and also helps to
avoid the convergence problems that can occur in extreme examples of
unrepresentative priors. More generally, this
approach highlights the intrinsic degeneracy between the `effective'
likelihood and prior in the formulation of Bayesian inference
problems, which it may prove advantageous to exploit using NS
methods more broadly than in merely addressing the unrepresentative
prior problem, although we will defer such considerations to
future publications. \revised{More discussion about generalised Bayesian prior design is given in \cite{simpson2017penalising}.}

This paper is organized as follows. Section~\ref{Sec:NestedSampling}
gives a brief summary of NS. Section
\ref{Sec:IncorrectPrior} details the underlying problem, and
illustrates it using a simple toy example. Section \ref{Sec:PSMethod}
describes the PR method and its implementation in the widely-used
NS algorithm MultiNest. Section \ref{Sec:NumEval} shows
some numerical results in simple synthetic examples. Section
\ref{Sec:Con} \revised{concludes the proposed approach and discusses its advantages and limitations}.

\section{Nested sampling}\label{Sec:NestedSampling}

NS is a sequential sampling method that can efficiently
explore the posterior distribution by repeatedly finding a higher
likelihood region while keeping the number of samples the same. It
consists of the following steps:
\begin{itemize}
\item A certain number \revised{($N_{\rm live}$) of samples of the parameters $\mtheta$ are drawn from the prior distribution $\mpi(\mtheta)$; these are termed `live points'}.
\item The likelihoods of these samples are computed through the
  likelihood function $\mL(\mtheta)$.
\item The sample with the lowest likelihood is removed and replaced by
  a sample again drawn from the prior, but constrained to a higher
  likelihood than that of the discarded sample.
\item The above step is repeated until some convergence criteria are
  met (e.g. the difference in evidence estimates between two
  iterations falls below a pre-defined threshold); the final set of
  samples and the discarded samples are then used to estimate the
  evidence $\mZ$ in model selection and obtain
  posterior-weighted samples for use in parameter estimation.
\end{itemize}

\revised{Pseudo code for the NS algorithm is given below.  Note that it is only one of the various possible NS implementations. Other implementations share the same structure but may differ in details, for example in how $X_i$ or $w_i$ is calculated, or the method used for drawing new samples. See \cite{skilling2006nested} for details.}

\begin{algorithm}[!ht]
\tcp{Nested sampling initialization}
At iteration $i=0$, draw $ N_{\rm live}$ samples $\{\mtheta_n\}_{n=1}^{N_{\rm live}}$ from prior $\pi(\mtheta)$ within prior space $\Psi$. Initialise evidence $Z=0$ and prior volume $X_0 = 1$. \\
\tcp{NS iterations}
\For{$i=1, 2, \cdots, I$}{
	$\bullet$ Compute likelihood $\mL(\mtheta_{n})$ for all $N_{\rm live}$ samples. \\
	$\bullet$ Find the lowest likelihood in live sample and save it in $\mL_i$. \\
	$\bullet$ Calculate weight $w_i = \frac{1}{2}(X_{i-1} - X_{i+1})$, where the prior volume $X_{i} = \exp(-i/N_{\rm live})$. \\
	$\bullet$ Increment evidence $Z$ by $\mL_i w_i$. \\
	$\bullet$ Replace the individual sample with likelihood $\mL_i$ by a newly drawn sample from restricted prior space $\Psi_i$ such that $\mtheta \in \Psi_i$ satisfies $\mL(\mtheta) > \mL_i$.\\ 
    $\bullet$ If $\mbox{max}\{\mL(\mtheta_n)\}X_i < \exp({\tt{tol}})Z$, then \bf{stop}. 
	}

Increment $Z$ by $\sum_{n=1}^{N_{\rm live}}\mL(\mtheta_n)X_{I}  / N_{\rm live}$. \\
Assign the sample replaced at iteration $i$ the importance weight $p_i=L_iw_i/Z$.\\
\caption{\revised{Nested sampling algorithm}}
\label{alg:NS}
\end{algorithm}

\revised{In Algorithm \ref{alg:NS}, $X_0$ represents the whole prior volume of prior space $\Psi$, and $\{X_i\}_{i=1}^I$ are the constrained prior volumes at each iteration. The number of iterations $I$ depends on a pre-defined convergence criterion {\tt{tol}} on the accuracy of the final log-evidence value and on the complexity of the problem.}

Among the various implementations of the NS algorithm,
two widely used packages are MultiNest
\citep{feroz2009multinest,feroz2013importance} and PolyChord
\citep{handley2015polychord}. MultiNest draws the new sample at each
iteration using rejection sampling from within a
multi-ellipsoid bound approximation to the iso-likelihood surface
defined by the discarded point; the bound is constructed from the
samples present at that iteration. PolyChord draws the new sample at
each iteration using a number of successive slice-sampling steps taken
in random directions. Please see \cite{feroz2009multinest} and
\cite{handley2015polychord} for more details.

\section{Unrepresentative prior problem} \label{Sec:IncorrectPrior}

We describe a prior $\mpi(\mtheta)$ as unrepresentative in the
analysis of a particular dataset, if the true values of the parameters
[or, more precisely, the peak(s) of the
  likelihood $\mL(\mtheta)$] for that dataset lie very far into the
wings of $\mpi(\mtheta)$. In real-world applications, this can occur
for a number of reasons, for example: (i) limited prior knowledge may
be available, resulting in a simple tractable distribution being
chosen as the prior, which could be unrepresentative; (ii) one may
wish to adopt the same prior across a large number of datasets that
might correspond to different true values of the parameters of
interest, and for some of these datasets the prior may be
unrepresentative. In any case, as we illustrate below in a simple
example, an unrepresentative prior may result in very inefficient
exploration of the parameter space, or failure of the sampling
algorithm to converge correctly in extreme cases. This can be
particularly damaging in applications where one wishes to perform
analyses on many thousands (or even millions) of different datasets,
since those (typically few) datasets for which the prior is
unrepresentative can absorb a large fraction of the computational
resources. Indeed, the authors have observed this phenomenon
in practice in an industrial geophysical application consisting of
only $\sim 1000$ different datasets.

It is also worth mentioning that one could, of course, encounter the
even more extreme case where the true parameter values, or likelihood
peak(s), for some dataset(s) lie outside an assumed prior having
compact support. This case, which one might describe as an {\em
  unsuitable} prior, is not addressed by our PR method, and is not
considered here.

\subsection{A univariate toy example}
\label{Sec:simpExp}
One may demonstrate the unrepresentative prior problem using a simple
one-dimensional toy example. Suppose one makes $N$ independent
measurements (or observations) $X = [x_{1}, \cdots, x_{n}, \cdots, x_{N}]^{\top}$ of some
quantity $\mtheta$, such that
\begin{align}
x_{n} = \mtheta + \mxi, \label{Eq:Simpexp}
\end{align}
where $\mxi$ denotes the simulated measurement noise, which is
Gaussian distributed $\mxi \sim \calN (\mu_{\mxi}, \sigma_{\mxi}^2)$
with mean $\mu_{\mxi}$ and variance $\sigma_{\mxi}^2$. For simplicity,
we will assume the measurement process is unbiased, so that
$\mu_{\mxi} = 0$, and that the variance $\sigma_{\mxi}^2$ of the noise
is known {\em a priori} (although it is a simple matter to relax these two
assumptions).

The likelihood $\mL(\mtheta)$ is therefore simply the product of $N$
Gaussian densities:
\begin{equation}
\mL(\mtheta) = \prod_{n=1}^{N} 
\left\{ \frac{1}{\sqrt{2 \pi \sigma_{\mxi}^2}} \exp \left[-\frac{(\mtheta - x_n)^2}{2 \sigma_{\mxi}^2}\right] \right\}.
\label{eqn:likelihood}
\end{equation}
For the purposes of illustration, we will assume the prior
$\mpi(\mtheta)$ also to be a Gaussian, with mean $\mu_{\mpi} = 0$ and
standard deviation $\sigma_{\mpi}=4$, such that {\em a priori} one
expects $\theta$ to lie in the range $[-10,10]$ with probability of
approximately $0.99$. Since the likelihood and prior are both
Gaussian in $\mtheta$, then so too is the posterior $\mP(\mtheta)$.

\begin{figure*}[!ht]
\centering
\subfigure[Case (1): $\mtheta_\ast = 5$]{
\includegraphics[width = 0.5\linewidth]{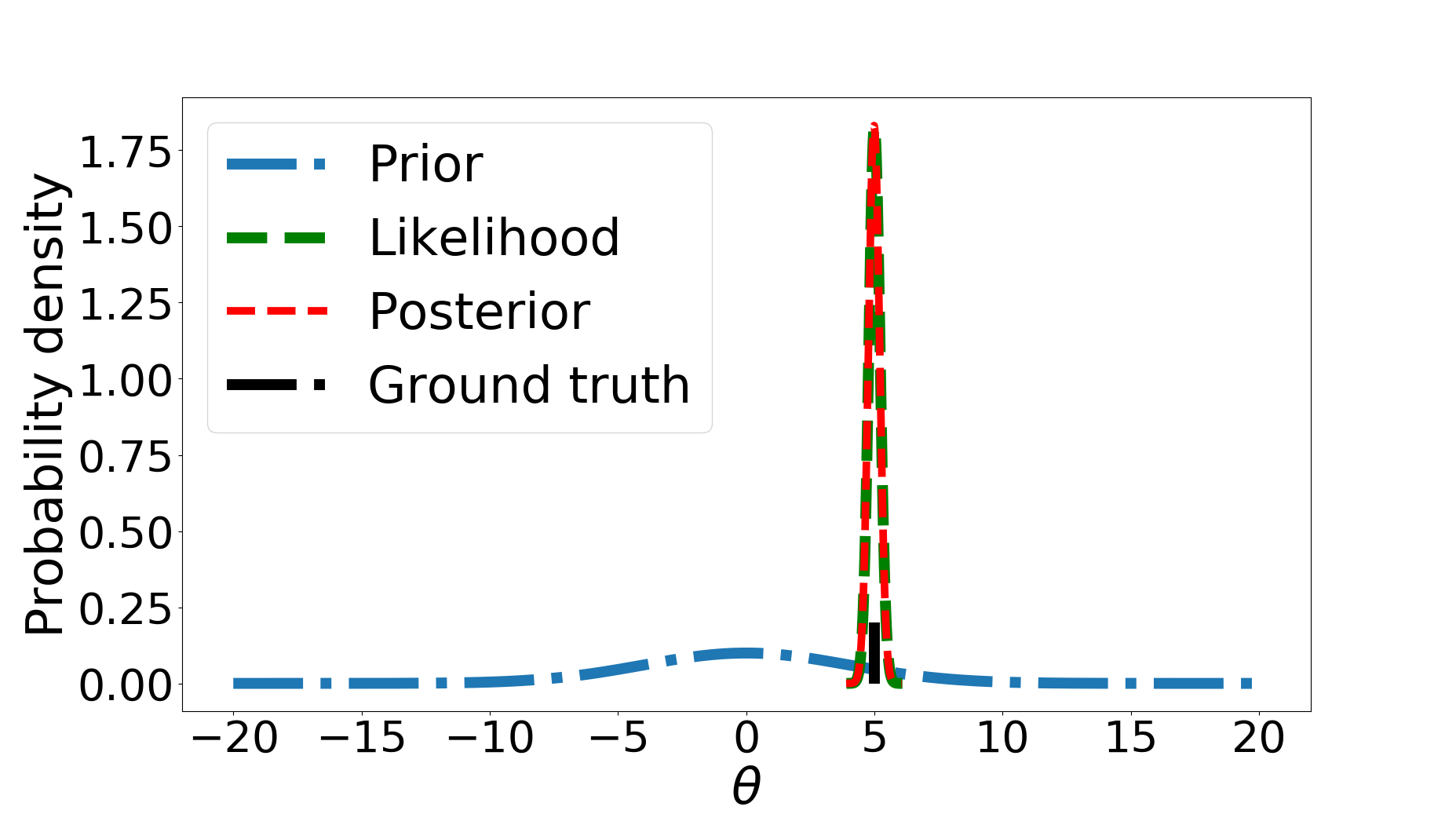}}
\subfigure[Case (1) estimation]{
\includegraphics[width = 0.46\linewidth]{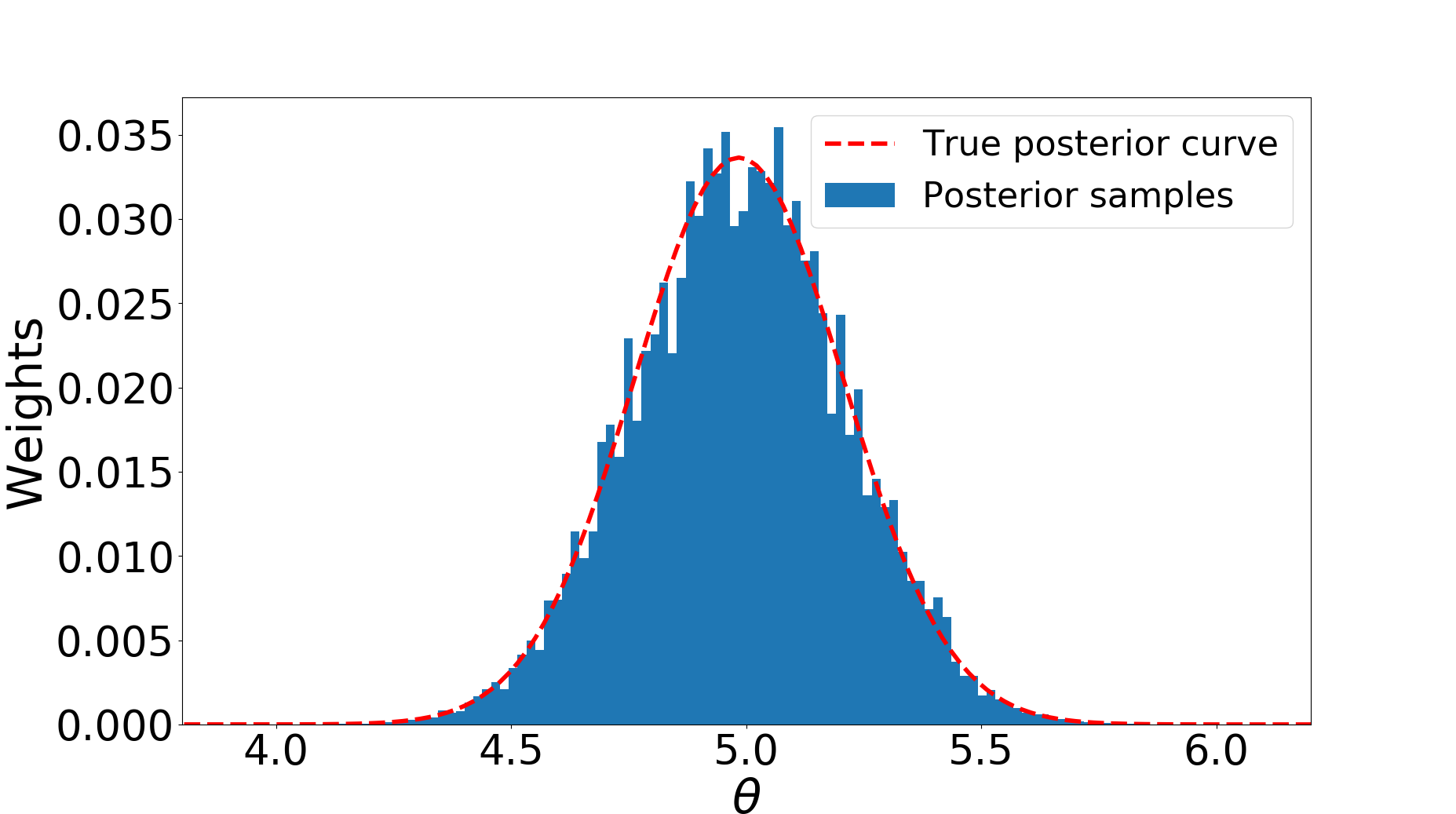}}
\vspace{3mm}
\subfigure[Case (2): $\mtheta_\ast = 30$]{
\includegraphics[width = 0.5\linewidth]{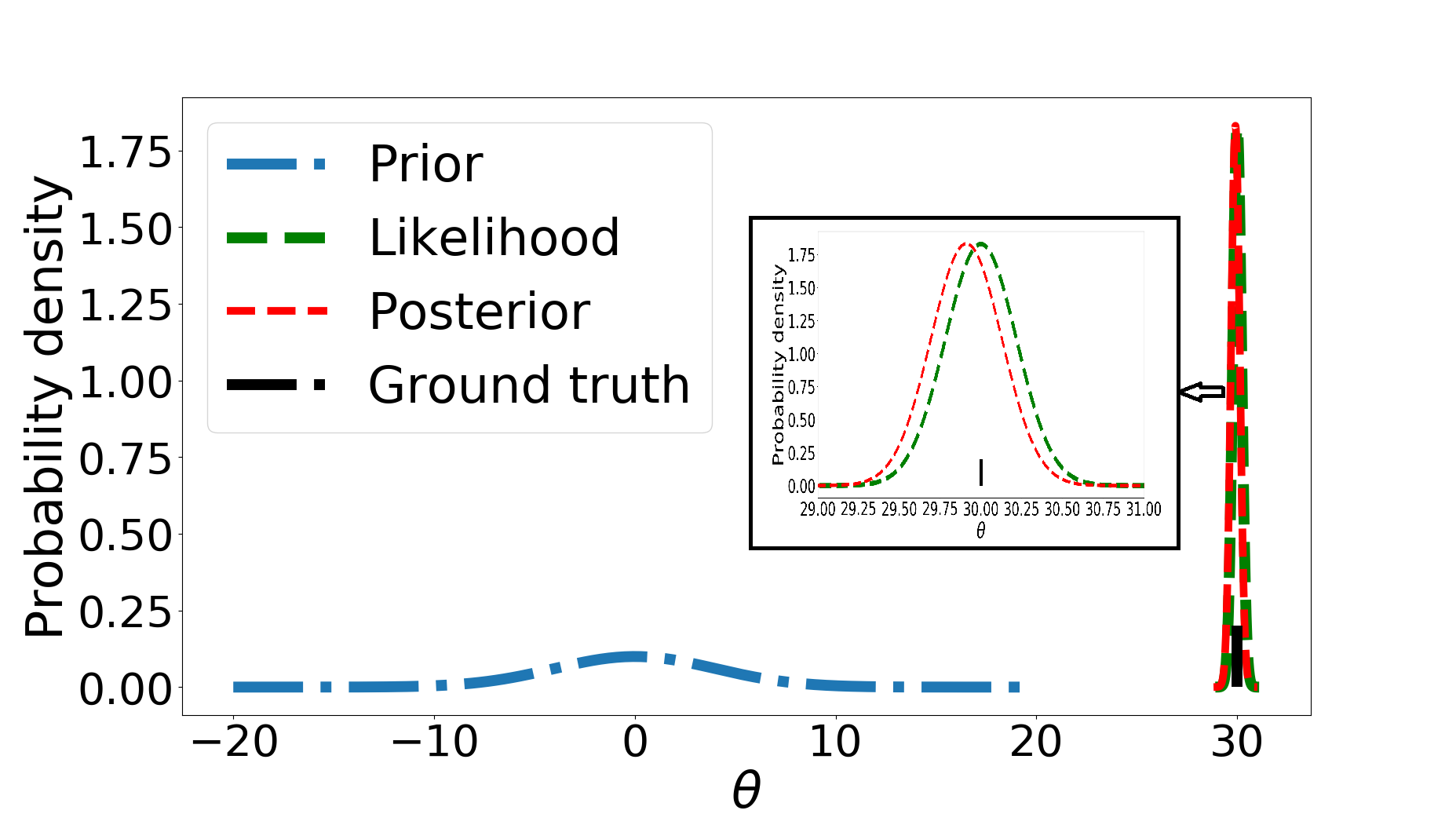}}
\subfigure[Case (2) estimation]{
\includegraphics[width = 0.46\linewidth]{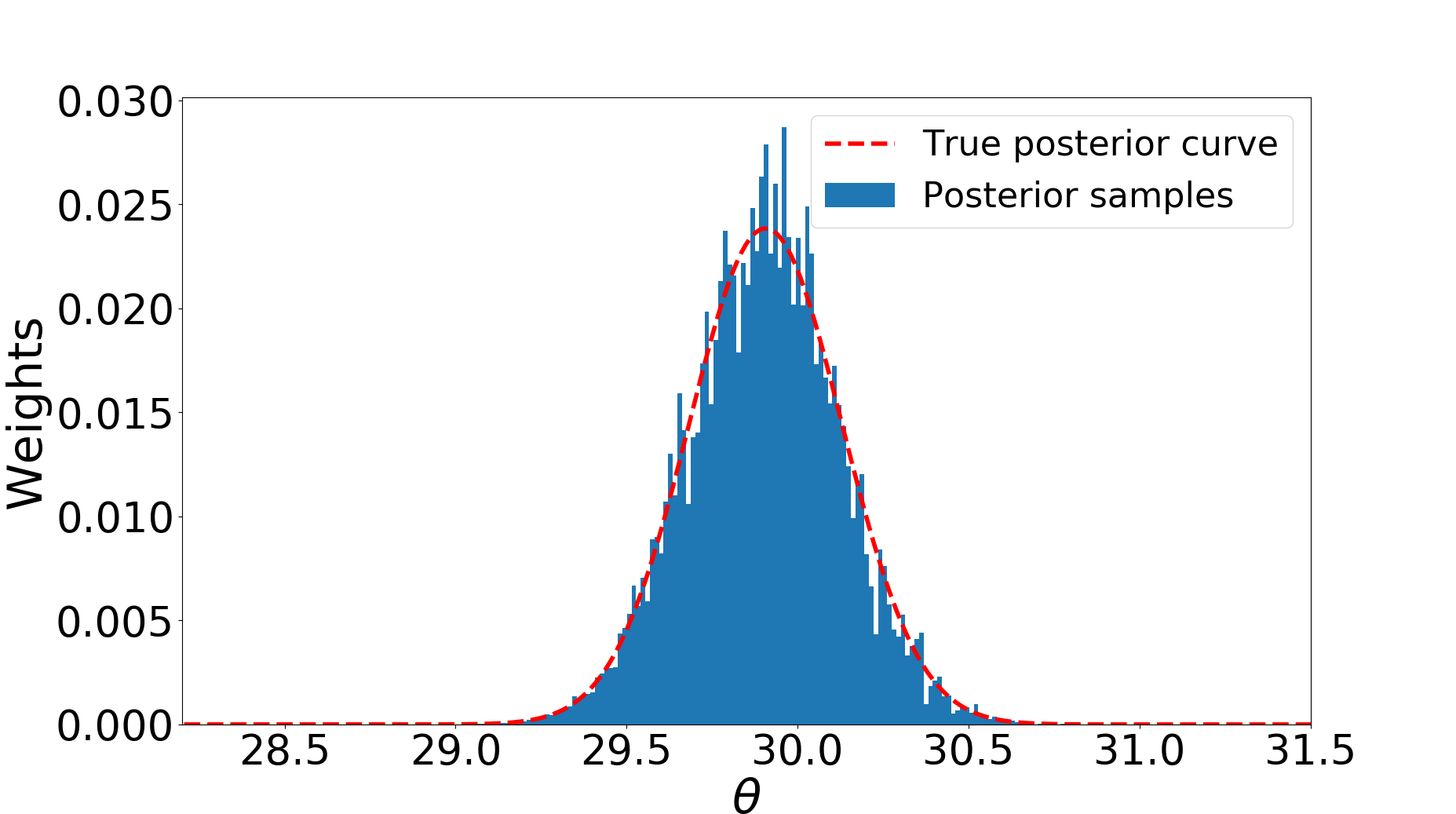}}
\vspace{3mm}
\subfigure[Case (3): $\mtheta_\ast = 40$]{
\includegraphics[width = 0.5\linewidth]{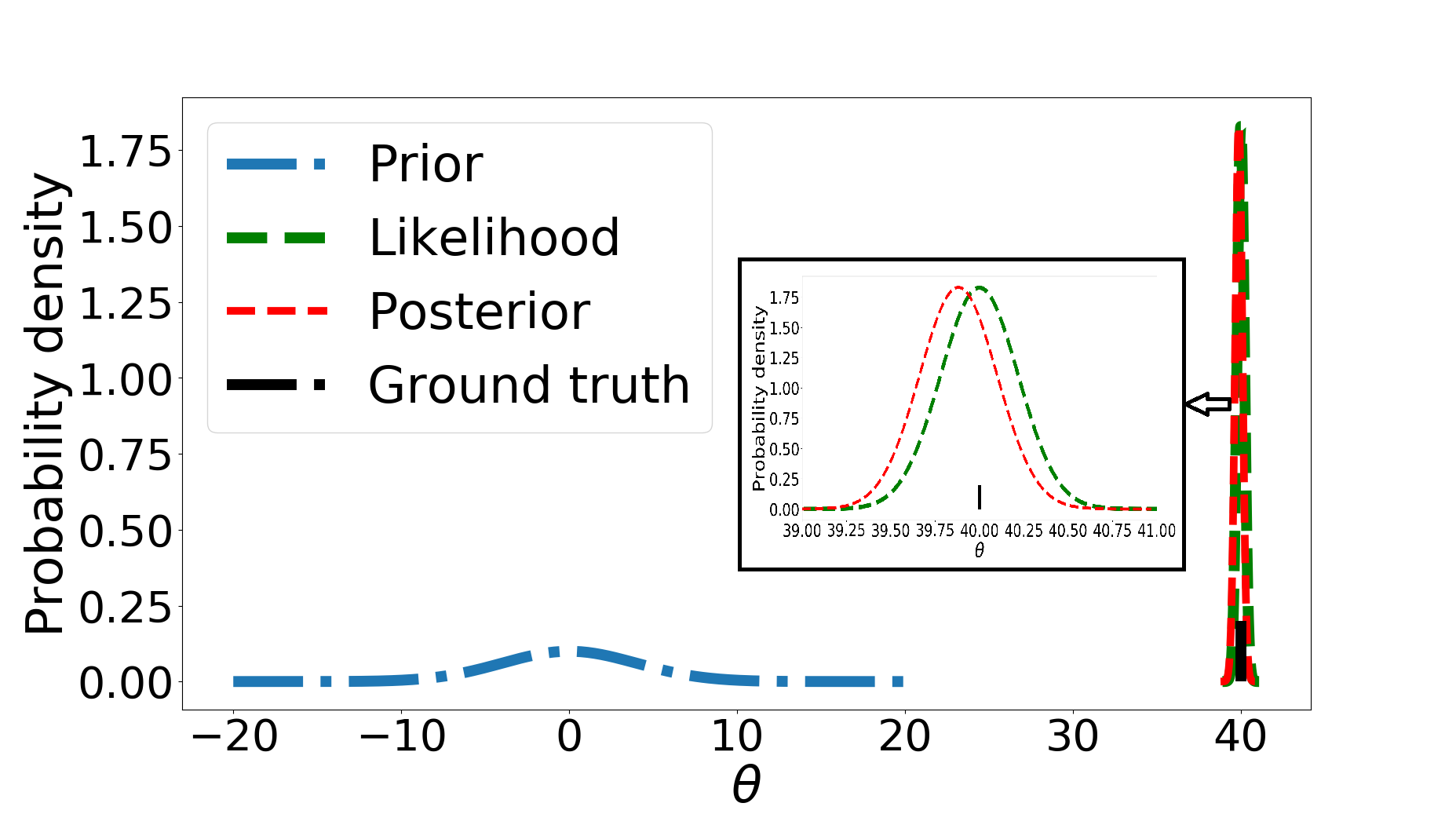}}
\subfigure[Case (3) estimation]{
\includegraphics[width = 0.46\linewidth]{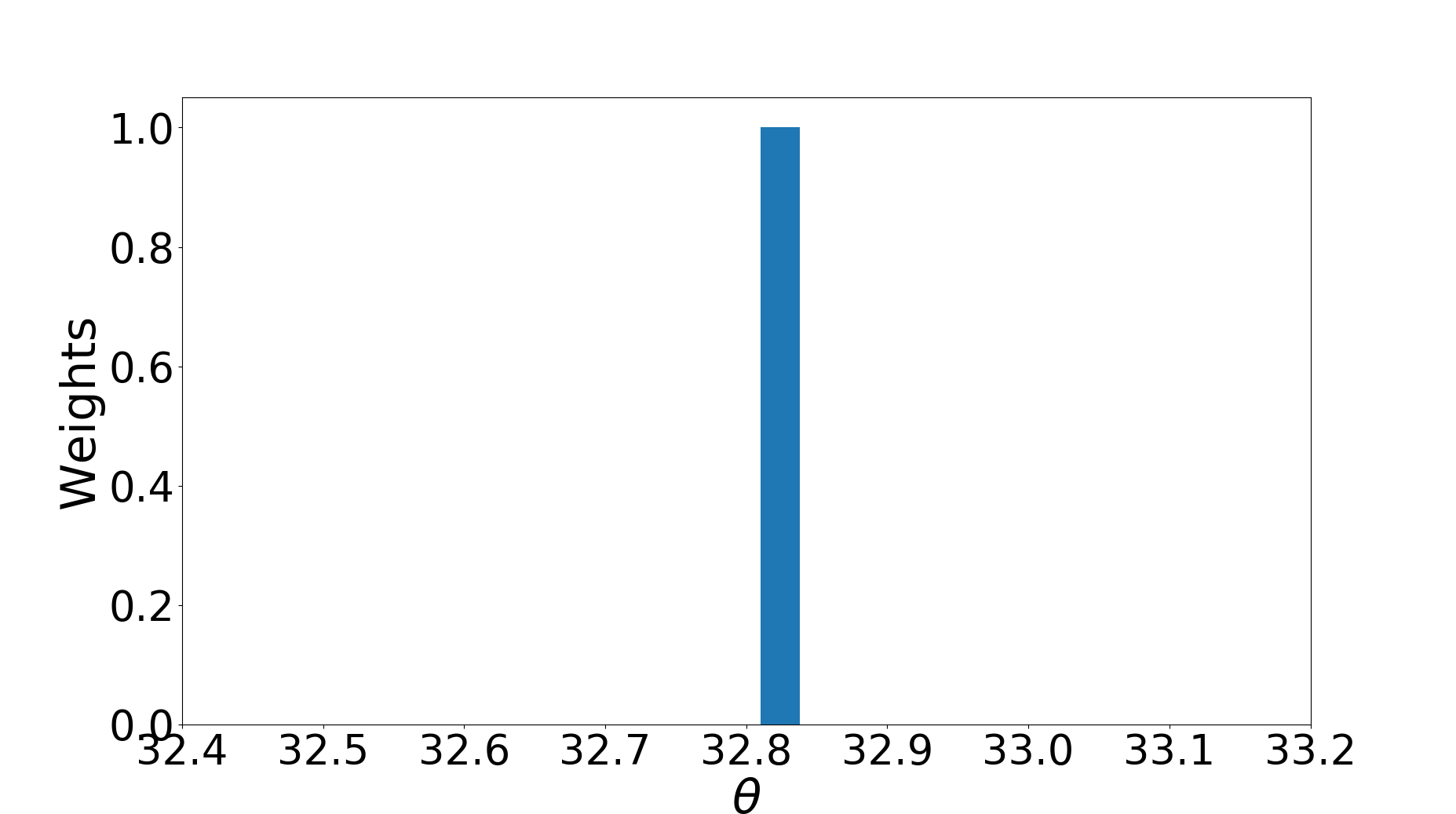}}
\caption{A univariate toy example illustrating the unrepresentative prior
  problem.  Sub-figures (a), (c) and (e) show, respectively, the cases
  (1), (2) and (3) discussed in the text; sub-figures (c) and (e)
  contain zoomed-in plots. The truth $\mtheta_\ast$ in each case is
  $\mtheta_\ast=5$, $\mtheta_\ast=30$ and $\mtheta_\ast=40$,
  respectively (dashed black lines). The prior (dashed blue curves) is
  a Gaussian distribution with $\mu_{\mpi}=0$ and $\sigma_{\mpi} =
  4$. The likelihood (dashed green curves) is a Gaussian
  (\ref{eqn:likelihood}) with $\mu_{\mxi}=1$.  According to Bayes
  theorem \eqref{Eq:Bayes}, the posterior (dashed red curves) is also
  a Gaussian calculated from the product of prior and
  likelihood. Sub-figures (b), (d) and (f) show, for each case, the
  histogram (blue bins) of posterior samples from MultiNest,
  and the true posterior distribution (solid red curves).}
\label{fig:prior1}
\end{figure*}

To illustrate the problem of an unrepresentative prior, we consider
three cases in which the true value $\mtheta_\ast$ of the unknown
parameter is given, respectively, by: (1) $\mtheta_\ast=5$, (2)
$\mtheta_\ast=30$ and (3) $\mtheta_\ast=40$. Thus, case (1)
corresponds to a straightforward situation in which the true value
$\mtheta_\ast$ lies comfortably within the prior, whereas cases (2)
and (3) represent the more unusual eventuality in which the true value
lies well into the wings of the prior distribution. In our simple
synthetic example, one expects cases (2) and (3) to occur only
extremely rarely. In real-world applications, however, the prior
distribution is typically constructed on a case-by-case basis by
analysts, and may not necessarily support a standard frequentist's
interpretation of the probability of `extreme' events. In fact, such
situations are regularly encountered in real-world applications, when
a large number of datasets are analysed. In each of the three cases
considered, we set the variance of the simulated measurement noise to
be $\sigma_{\mxi} = 1$ and the number of measurements is $N=20$. Note
that the width of the likelihood in \eqref{eqn:likelihood} is
proportional to $1/\sqrt{N}$, so the unrepresentative prior problem
becomes more acute as $N$ increases.

Figures \ref{fig:prior1} (a), (c) and (e) show the prior, likelihood
and posterior distributions for the cases (1), (2) and (3),
respectively. One sees that as the true value $\mtheta_\ast$ increases
and lies further into the wings of the prior, the posterior lies
progressively further to the left of the likelihood, as expected.  As
a result, in cases (2) and (3), the peak of the posterior (red dashed
curve) is displaced to the left of the true value (black dashed
line). This can be clearly observed in the zoomed-in plots within
sub-figures (c) and (e).  Figures~\ref{fig:prior1} (b), (d) and
(f) show histograms (blue bins) of the posterior samples obtained
using MultiNest for cases (1), (2) and (3), respectively, together
with the corresponding true analytical posterior distributions (red
solid curves). In each case, the MultiNest sampling parameters were
set to $N_{\rm live} = 2000$, ${\rm efr} = 0.8$ and ${\rm tol} = 0.5$
(see \citealt{feroz2009multinest} for details), and the algorithm was
run to convergence. A natural estimator $\hat{\mtheta}$ and
uncertainty $\Delta\mtheta$, respectively, for the value of the
unknown parameter are provided by the mean and standard deviation of
the posterior samples in each case, and are given in
Table~\ref{tab:example}.

In case (1), one sees that the samples obtained are indeed consistent
with being drawn from the true posterior, as expected.  The mean
$\hat{\mtheta}$ and standard deviation $\Delta\mtheta$ of the samples
listed in Table~\ref{tab:example} agree well with the mean
$\mu_\mathcal{P}$ and standard deviation $\sigma_\mathcal{P}$ of the
true posterior distribution. In this case, MultiNest converged
relatively quickly, requiring a total of $13529$ likelihood
evaluations. On repeating the entire analysis a total of 10 times, one
obtains statistically consistent results in each case.

In case (2), one sees that the samples obtained are again consistent
with being drawn from the true posterior. Indeed, from
Table~\ref{tab:example}, one may verify that the mean 
and standard deviation of the samples agree well
with those of the true posterior distribution. In this case, however,
the convergence of MultiNest is much slower, requiring about 6 times
the number of likelihood evaluations needed in case (1). This is a
result of the true value lying far out in the wings of the prior
distribution. Recall that NS begins by drawing $N_{\rm
  live}$ samples from the prior and at each subsequent iteration
replaces the sample having the lowest likelihood with a sample again
drawn from the prior but constrained to have a higher
likelihood. Thus, as the iterations progress, the collection of
$N_{\rm live}$ `live points' gradually migrates from the prior to the
peak of the likelihood. When the likelihood is concentrated very far out in the wings of the prior, this process can become very slow, even if one is able to draw
each new sample from the constrained prior using standard methods
(sometimes termed perfect nested sampling). In practice, this is
usually not possible, so algorithms such as MultiNest and PolyChord use other methods that may require several likelihood evaluations
before a new sample is accepted. Depending on the method used, an
unrepresentative prior can also result in a significant drop in
sampling efficiency, thereby increasing the required number of
likelihood evaluations still further. On repeating the entire analysis
a total of 10 times, once again obtains statistically consistent
results in each case.

In case (3), one sees that the samples obtained are clearly
inconsistent with being drawn from the true posterior. Indeed, the
samples are concentrated at just a single value of $\mtheta$.  This
behaviour may be understood by again considering the operation of
NS. The algorithm begins by drawing $N_{\rm live}=2000$ samples from
the prior, which is a Gaussian with mean $\mu_{\mpi} = 0$ and standard
deviation $\sigma_{\mpi}=4$. Thus, one would expect approximately only
one such sample to lie outside the range $[-14,14]$. \revised{Moreover, since the likelihood is a Gaussian centred near the true value $\mtheta_\ast=40$ with standard deviation $\sim 0.25$, the live points will typically all lie in a region over which the likelihood is very small and flat (although, in this particular example, the values of the log-likelihood for the live points -- which is the quantity used in the numerical calculations -- are still distinguishable to machine precision).}

\revised{When the point with the lowest likelihood value is discarded,
  it must be replaced at the next NS iteration by another drawn from
  the prior, but with a larger likelihood. How this replacement sample
  is obtained depends on the particular NS implementation being used.
  As discussed in Section~\ref{Sec:NestedSampling}, MultiNest draws
  candidate replacement samples at each iteration using rejection
  sampling from within a multi-ellipsoid bound approximation to the
  iso-likelihood surface defined by the discarded point, which in just
  one dimension reduces simply to a range in $\mtheta$. Since
  this bound is constructed from the samples present at that
  iteration, it will typically not extend far beyond the locations of
  the live points having the extreme values of the parameter
  $\mtheta$. Thus, there is very limited opportunity to sample
  candidate replacement points from much larger values of $\mtheta$,
  where the likelihood is significantly higher. Hence, as the NS
  iterations proceed, the migration of points from the prior towards
  the likelihood is extremely slow. Indeed, in this case, the
  migration is sufficiently slow that the algorithm terminates (in
  this case after $96512$ likelihood evaluations) before reaching the
  main body of the likelihood and produces a set of posterior-weighted
  samples from the discarded points (see \citealt{feroz2009multinest}
  for details). Since this weighting is proportional to the
  likelihood, in this extreme case the recovered posterior is merely a
  `spike' corresponding to the sample with the largest likelihood, as
  observed in Figure~\ref{fig:prior1} (f). In short, the algorithm has
  catastrophically failed.  On repeating the entire analysis a total
  of 10 times, one finds similar pathological behaviour in each case.}

\revised{One may, of course, seek to improve the performance of NS in
  such cases in a number of ways. Firstly, one may adjust the
  convergence criterion ({\tt{tol}} in MultiNest) so that many
  more NS iterations are performed, although there is no guarantee in
  any given problem that this will be sufficient to prevent premature
  convergence. Perhaps more useful is to ensure that there is a
  greater opportunity at each NS iteration of drawing candidate
  replacement points from larger values of $\mtheta$, where the
  likelihood is larger. This may be achieved in a variety of ways. In
  MultiNest, for example, one may reduce the $\tt{efr}$ parameter so
  that the volume of the ellipsoidal bound (or the $\mtheta$-range in
  this one-dimensional problem) becomes larger. Alternatively, as in
  other NS implementations, one may draw candidate replacement points
  using either MCMC sampling \citep{feroz2008multimodalns} or
  slice-sampling \citep{handley2015polychord} and increase the number
  of steps taken before a candidate point is chosen}.

\revised{All the of above approaches may mitigate the problem to some
  degree in particular cases (as we have verified in further numerical
  tests), but only at the cost of a simultaneous dramatic drop in
  sampling efficiency caused precisely by the changes made in
  obtaining candidate replacement points. Moreover, in more extreme
  cases these measures fail completely. In particular, if the
  prior and the likelihood are extremely widely separated, the
  differences in the values of the log-likelihood of the live samples
  may fall below the machine accuracy used to perform the
  calculations. Thus, the original set of prior-distributed samples
  are likely to have log-likelihood values that are indistinguishable
  to machine precision. Thus, the `lowest likelihood' sample to be
  discarded will be chosen effectively at random. Moreover, in seeking
  a replacement sample that is drawn from the prior but having a
  larger likelihood, the algorithm is very unlikely to obtain a sample
  for which the likelihood value is genuinely larger to machine
  precision. Even if such a sample is obtained, then the above
  problems will re-occur in the next iteration when seeking to replace
  the next discarded sample, and so on. In this scenario, the sampling
  efficiency again drops dramatically, but more importantly the
  algorithm essentially becomes stuck and will catastrophically fail
  because of accumulated numerical inaccuracies.}

\begin{table}
\caption{MultiNest performance in the toy example illustrated in
  Figure~\ref{fig:prior1}.}
\begin{tabular}{lccc}
\hline
 & Case (1) & Case (2) & Case (3)  \\
\hline
True value $\theta_\ast$ & 5 & 30 & 40 \\
True posterior $\mu_{\cal{P}}$ & 4.984 & 29.907 & 39.875 \\
True posterior $\sigma_{\cal{P}}$ &  0.223  &  0.223  &  0.223 \\
\hline
Likelihood calls & 13529 & 78877 & 96512 \\
Estimated value $\hat{\mtheta}$ & 4.981 & 29.902 & 32.838 \\ 
Uncertainty $\Delta\mtheta$ & 0.223 & 0.223 & \num{7.6e-6}  \\ 
\hline
\end{tabular}
\label{tab:example}
\end{table}

\subsection{Simple `solutions'}

A number of potential simple `solutions' to the unrepresentative prior
problem are immediately apparent. For example, one might consider the
following:
\begin{itemize}
\item modify the prior distribution across one's analysis, either by
  increasing its standard deviation $\sigma_{\pi}$, or even by
  adopting a different functional form, so that it should comfortably
  encompass the likelihood for all datasets;
\item perform the analysis using the original prior for all the
  datasets, identify the datasets for which it is unrepresentative by
  monitoring the sampling efficiency and examining the final set of
  posterior samples for pathologies, and then modify the prior as
  above for these datasets.
\end{itemize}

Unfortunately, neither of these approaches is appropriate or
realistic. The former approach is inapplicable since the prior may be
representative for the vast majority of the datasets under analysis,
and one should use this information in deriving inferences. Also, the
former solution sacrifices the overall speed and computational
efficiency, as the augmented prior is applied to all cases but not
only the problematic ones. Choosing a proper trade-off between the
efficiency and the coverage of prior is difficult when a large number
of experiments need to be examined.

The latter solution requires one to identify various outlier cases (as
the outlier cases could be very different from one to another), and
also perform re-runs of those identified. It becomes a non-trivial
computational problem when a single algorithm run requires a
considerable amount of run time, or when the results of the outlier
cases are needed for the next step computation, i.e. the whole process
waits for the outlier cases to proceed.  This could be trivial
for some applications and could be very difficult for others in which
many different outlier cases exist.

\section{Posterior repartitioning method} \label{Sec:PSMethod}

The posterior repartitioning (PR) method addresses the unrepresentative
prior problem in the context of NS algorithms
\citep{skilling2006nested} for exploring the parameter space, without
sacrificing computational speed or changing the inference obtained.

\subsection{General expressions}

\revised{In general, the `repartition' of the product $\mL(\mtheta)\pi(\mtheta)$ can be expressed as:
\begin{align}
\mL(\mtheta) \mpi(\mtheta) = \tmL(\mtheta) \tmpi(\mtheta),
\label{Eq:equality}
\end{align}
where $\tmL(\mtheta)$ and $\tmpi(\mtheta)$ are the new effectivelikelihood and prior, respectively. As a result, the (unnormalised) posterior remains unchanged.} The \textit{modified prior} $\tmpi(\mtheta)$ can be any tractable distribution, which we assume to be appropriately normalised to unit volume. \revised{The possibility of repartitioning the posterior in NS was first mentioned in \cite{feroz2009multinest}, but equation (\ref{Eq:equality}) can also be viewed as the vanilla case (when the importance weight function equals to $1$) of nested importance sampling proposed in \cite{chopin2010properties}.}

One general advantage of NS is that the evidence (or
marginal likelihood), which is intractable in most cases, can be
accurately approximated. This is achieved by first defining $\mV(\ml)$
as the prior volume within the iso-likelihood contour $\mL(\mtheta) =
\ml$, namely
\begin{align}
\mV(\ml) = \int_{\mL(\mtheta)> \ml} \mpi(\mtheta) d \mtheta,
\end{align}
where $\ml$ is a real number that gradually rises from zero to the
maximum of $\mL(\mtheta)$ as the NS iterations progress,
so that $\mV(\ml)$ monotonically decreases from unity to zero. After PR, $\mpi(\mtheta)$ is replaced by $\tmpi(\mtheta)$, and the evidence can be calculated as
\begin{align}
\mZ = \int \mL(\mtheta) \mpi(\mtheta) d \mtheta 
= \int \tmL(\mtheta) \tmpi(\mtheta) d \mtheta=
\int_{0}^{1} \mL(\mV) d \mV.
\end{align}
It is worth noting, however, that in the case where 
$\tmpi(\mtheta)$ is not properly normalised, \revised{the `modified evidence' $\mZ^\prime$} obtained after PR is simply related to the original evidence by
\begin{align}
\mZ  = \mZ^\prime \int \tmpi(\mtheta) d \mtheta. \label{Eq:EvitoEvi}
\end{align}
\revised{Provided one can evaluate the volume of the modified prior $\tmpi(\mtheta)$, one may therefore straightforwardly recover the original evidence, if required. For many simple choices of $\tmpi(\mtheta)$, this is possible analytically, but may require numerical integration in general. It should be noted, however, that the normalistion of the modified prior is irrelevant for obtaining posterior samples.} We now discuss some particular special choices for $\tmpi(\mtheta)$.

\subsection{Power posterior repartitioning}

Rather than introducing a completely new prior distribution into the
problem, a sensible choice is often simply to take
$\tmpi(\mtheta)$ to be the original prior $\mpi(\mtheta)$ raised to
some power, and then renormalised to unit volume, such that
\begin{align}
\tmpi(\mtheta) &= \frac{\mpi(\mtheta)^{\mbeta}}{\mZ_\pi(\beta)},
\\ \label{Eq:proportion} 
\tmL(\mtheta) &= \mL(\mtheta)
\mpi(\mtheta)^{(1-\mbeta)}\mZ_\pi(\beta),
\end{align}  
where $\mbeta \in [0,1]$ and $\mZ_\pi(\beta) \equiv \int
\mpi(\mtheta)^{\mbeta} d \mtheta$. 
By altering the value of $\mbeta$, the modified prior can be chosen
from a range between the original prior ($\mbeta=1$) and the uniform
distribution ($\mbeta=0$). As long as the equality in equation
\eqref{Eq:equality} holds, the PR method can be applied separately for
multiple unknown parameters with different forms of prior
distributions.

\begin{figure}[t]
\centering \includegraphics[width = 0.95\linewidth]{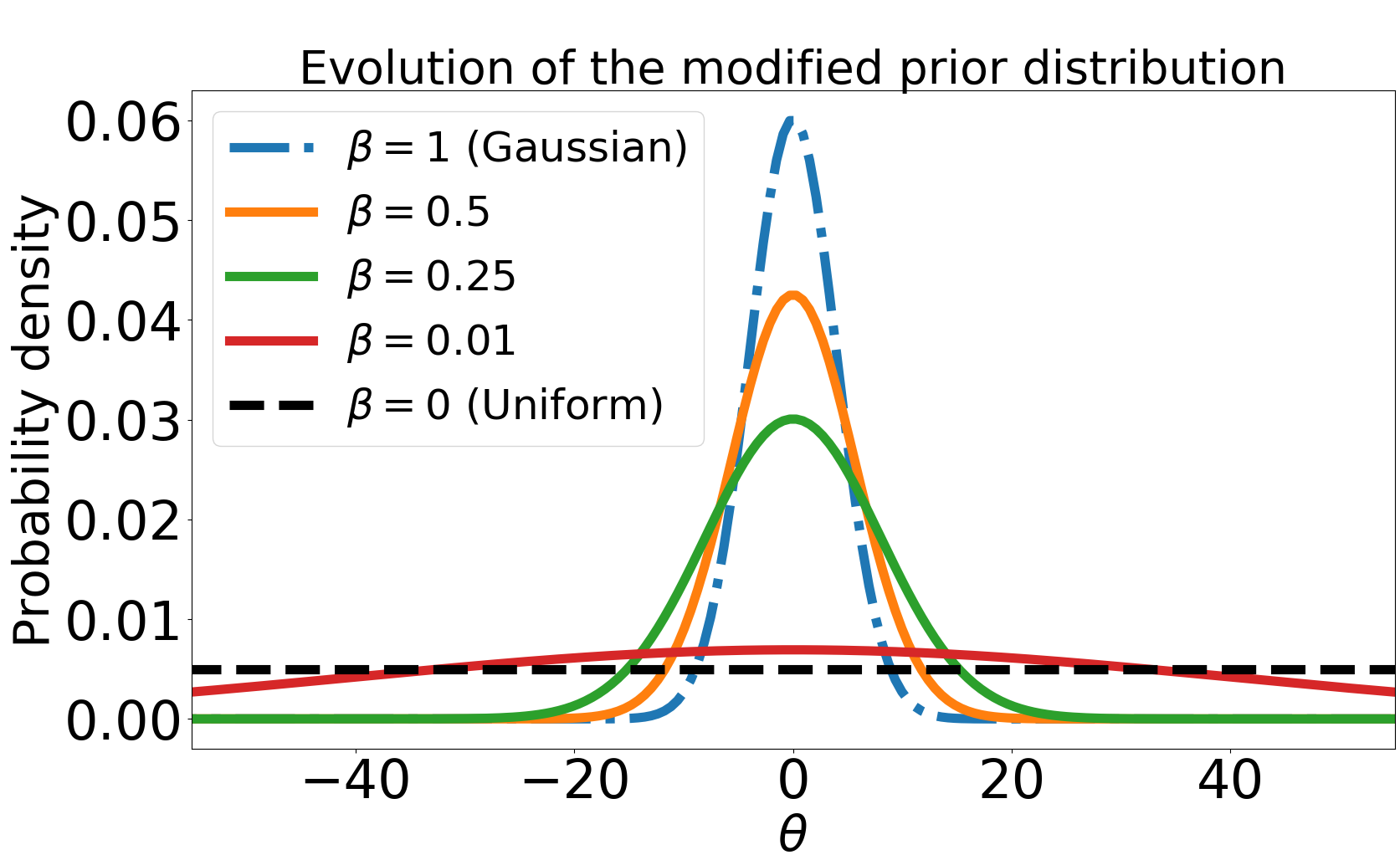}
\caption{One dimensional prior evolution for $\mbeta \in [0,1]$. The
  original prior is a Gaussian distribution with $\sigma_\pi=4$
  (truncated in the range $[-50,50]$) when $\mbeta=1$ (dashed blue
  curve), and is an uniform distribution when $\mbeta=0$ (dashed black
  curve). The remaining three curves correspond to $\mbeta=0.5$ (green
  curve), $0.25$ (red curve), $0.01$ (light blue curve),
  respectively.}
\label{fig:priorEvo}
\end{figure}

Figure \ref{fig:priorEvo} illustrates how the prior changes for
different values of $\mbeta$ in a one-dimensional problem.  As the
parameter $\mbeta$ decreases from $1$ to $0$, the prior distribution
evolves from a Gaussian centred on zero with standard deviation
$\sigma_\pi=4$ to a uniform distribution, where the normalisation
depends on the assumed support $[-50,50]$ of the unknown parameter
$\mtheta$. Indeed, the uniform modified prior $\tmpi(\mtheta) \sim
\calU(a, b)$ is a special case, but often a useful choice.  One
advantage of this choice is that the range $[a,b]$ can be easily set
such that it accommodates the range of $\mtheta$ values required to
overcome the unrepresentative prior problem, and the modified prior is
trivially normalised. It can cause the sampling to be inefficient,
however, since it essentially maximally broadens the search space
(within the desired range).

\label{Sec:powerPR}

The above approach is easily extended to multivariate problems with
parameter vector $\mTheta = (\mtheta_1, \mtheta_2, \cdots,
  \mtheta_N)^{\rm T}$. It is worth noting in particular the case where the
original prior is a multivariate Gaussian, \revised{such that $\mpi(\mTheta)
= \calN(\mmu, \mSigma)$}, where $\mmu$ is the vector of means for
each variable and $\mSigma$ is the covariance matrix. \revised{The power
modified prior $\tmpi(\mTheta)$ is then given simply by $\calN(\mmu, \mbeta^{-1}\mSigma)$} over the assumed supported region $\mathcal{R}$
of the parameter space,
and 
\begin{align}
\mZ_\pi(\beta) = 
(2 \mpi)^{\frac{\mN}{2}(1-\mbeta)} |\mSigma|^{\frac{(1-\mbeta)}{2}} \mbeta^{-\frac{\mN}{2}}
\int_{\mathcal{R}} \calN(\mmu,\mbeta^{-1}\mSigma)d \mTheta.
\label{Eq:eviExpan}
\end{align}

\revised{There is unfortunately no robust universal guideline for
  choosing an appropriate value for $\mbeta$, since this depends on
  the dimensionality and complexity of the posterior and on the
  initial prior distribution assumed. Nonetheless, as demonstrated in
  the numerical examples presented in Section~\ref{Sec:NumEval}, there
  is a straightforward approach for employing the PR method in more
  realistic problems, in which the true posterior is not
  known. Namely, starting from $\mbeta=1$ (which corresponds to the
  original prior), one can obtain inferences for progressively smaller
  values of $\mbeta$, according to some pre-defined or dynamic
  `annealing schedule', until the results converge to a statistically
  consistent solution. The precise nature of the annealing schedule is
  unimportant, although either linearly or exponentially decreasing
  values of $\mbeta$ seem the most natural approaches.}

\subsection{More general posterior repartitioning}

Raising the original prior to some power $\beta$ merely provides a
convenient way of defining the modified prior, since it essentially
just broadens the original prior by some specified amount. In general,
however, $\tmpi(\mtheta)$ can be any tractable distribution. For
example, there is no requirement for the modified prior to be centred
at the same parameter value as the original prior. One could,
therefore, choose a modified prior that broadens and/or shifts the
original one, or a modified prior that has a different form from the
original. Note that, in this generalised setting, the modified prior
should at least be non-zero everywhere that the original prior is non-zero.

\subsection{Diagnostics of the unrepresentative prior problem}
\label{sec:diagnostics}

This paper focuses primarily on how to mitigate the
unrepresentative prior problem using PR.  Another
critical issue, however, is how one may determine when the prior is
unrepresentative in the course the analysis of some (large number of)
dataset(s). We comment briefly on this issue here.

\begin{figure}[!ht]
\centering
\includegraphics[width = 1\linewidth]{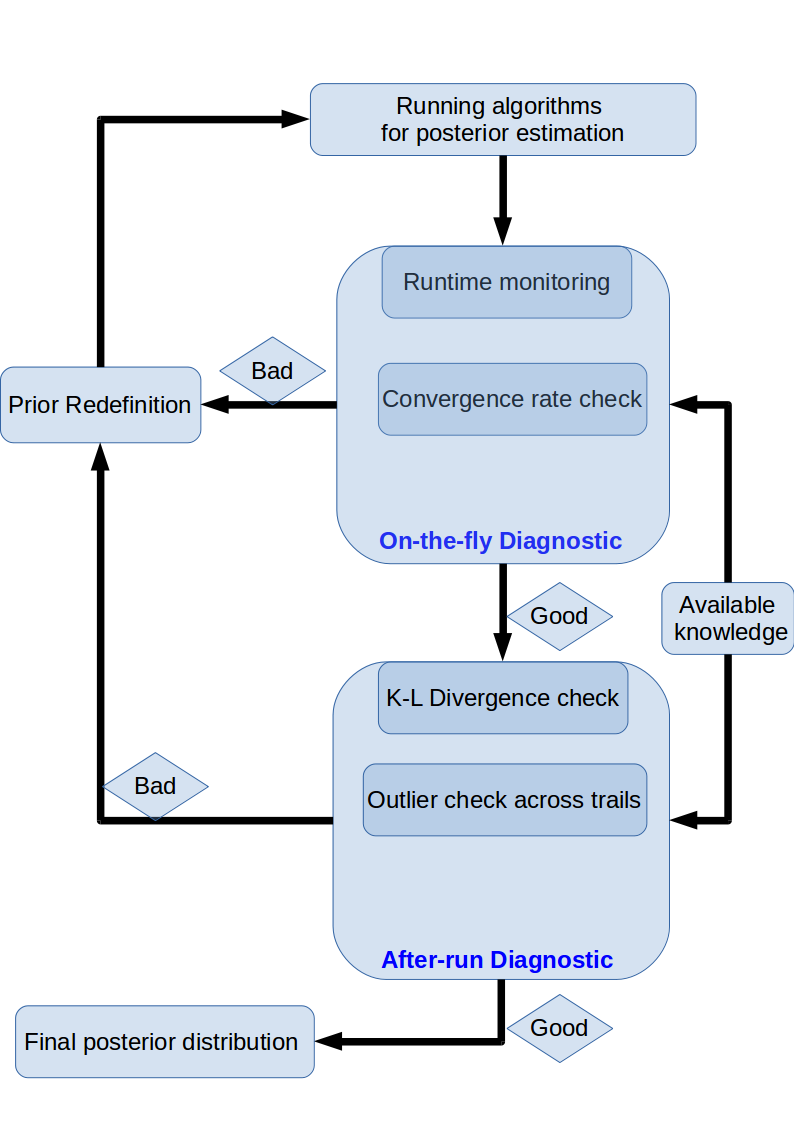}
\caption{A flow chart of a designed diagnostic process. The two main steps
  of the diagnostic process are highlighted in dark blue. The process
  starts by running a sampling algorithm for Bayesian parameter
  estimation (the top small block), and proceeds with two hierarchical
  diagnostics steps to evaluate the trail of interest. `Available
  knowledge' is defined as reliable experimental information and prior
  knowledge that one could obtain in advance.}
\label{fig:diagPic}
\end{figure}

Diagnosing the unrepresentative prior problem beforehand is generally
difficult. Thus, designing a practical engineering-oriented solution
is helpful in addressing most such problems. The goal of this
diagnostic is to identify abnormal cases amongst a number of datasets
during the analysis procedure. We assume that at least a few
`reliable' (sometimes called `gold standard') datasets, which do not
suffer from the unrepresentative prior problem, have been analysed
before the diagnostics. \revised{The reliability threshold of a dataset varies depending on different scenarios, but (ideally) a gold standard dataset should: (1) be recognised as such by field experts; (2) have all of its noise sources clearly identified and characterised; (3) yield parameter estimates that are consistent with true values either known a priori or determined by other means.} These provide us with some rough but reliable information and prior knowledge, such as runtime, convergence rate, and the shape of posterior distribution. We denote this information as the \textit{available knowledge} for the problem of interest.

One may then employ a diagnostic scheme of the type illustrated in
Figure \ref{fig:diagPic}, which is composed of two parts:
\textit{on-the-fly diagnostics} and \textit{after-run
  diagnostics}. On-the-fly diagnostics involve monitoring the runtime
and convergence status during the analysis of each dataset.
Specifically, runtime monitoring involves simply checking whether the
runtime of an individual analysis is greatly different from those of
the available knowledge. Similarly, convergence rate checks compare
the speed of convergence between the current run and the available
knowledge.  If both results are consistent with those in the available
knowledge, the diagnostic process proceeds to after-run diagnostics. \revised{Note that the quantitative consistency check can be defined in various ways. A simple method is to set a threshold for the difference between available knowledge and individual runs. For instance, the result from an individual run can be considered as a reliable one if the error between the individual run result and the mean of the available knowledge is within a certain threshold. Such criteria should be carefully discussed by field experts on a case-by-case basis.}

After-run diagnostics compare the computed posterior with the
available knowledge.  One plausible after-run diagnostic is to
evaluate some `distance' measure between the assumed prior and the
posterior distribution resulting from the analysis. An obvious choice
is to employ the Kullback--Leibler (KL) divergence (see, e.g.,
\citealt{bishop2006}). The KL divergence quantifies the difference
between two probability distributions by calculating their relative
entropy. A larger KL divergence indicates a greater difference between
the two distributions.  The KL divergence is, however, an asymmetric
measure and its value is not bounded. To overcome these drawbacks, one
could also consider the Jensen--Shannon divergence \citep{endres2003},
which is a symmetric variant of the KL divergence. The posterior may
also be compared with the available knowledge in the outlier check
step.

\revised{Finally, we note that a diagnostic analysis is valid when it is performed using the same algorithm specifications. For instance, $N_{\rm live}$, ${\rm efr}$, and ${\rm tol}$ settings should be the same in MultiNest when performing diagnostic analysis}. In any case, once a reasonable diagnostic metric is constructed, the abnormal trials can be identified according to some predetermined criteria and examined, and the proposed PR scheme can be applied on a case-by-case
basis.  A simple illustration of this process is presented in the bivariate example case in the next section.

\section{Numerical examples} \label{Sec:NumEval}

\revised{We begin by illustrating the PR method in two numerical examples, one
univariate and the other a bivariate Gaussian posterior.  Our
investigation is then extended to higher dimensional (from 3 to 15
dimensions) Gaussian posteriors, to explore its stability to the
`curse of dimensionality'.  Finally, we consider a bivariate
non-Gaussian example.} In particular, we compare the performance of the
MultiNest sampler before and after applying PR.

\begin{figure*}[!t]
\centering
\subfigure[$\mtheta_\ast = 40$, theoretical distribution]{
\includegraphics[width = 0.51\linewidth]{1DExp40}}
\hspace*{-0.2cm}\subfigure[$\mbeta = 0.8$]{
\includegraphics[width = 0.45\linewidth]{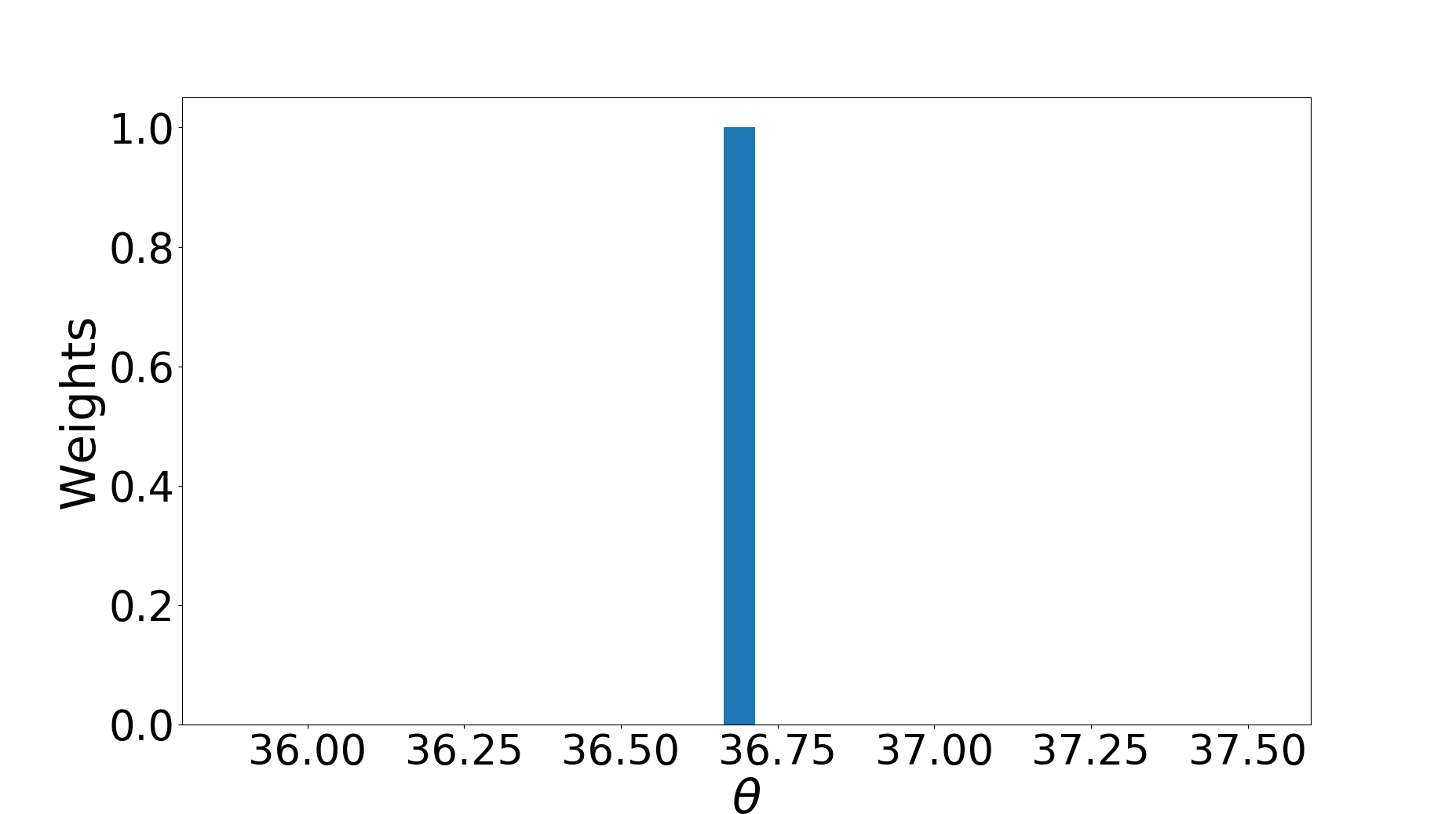}}
\subfigure[$\mbeta = 0.6$]{
\includegraphics[width = 0.46\linewidth]{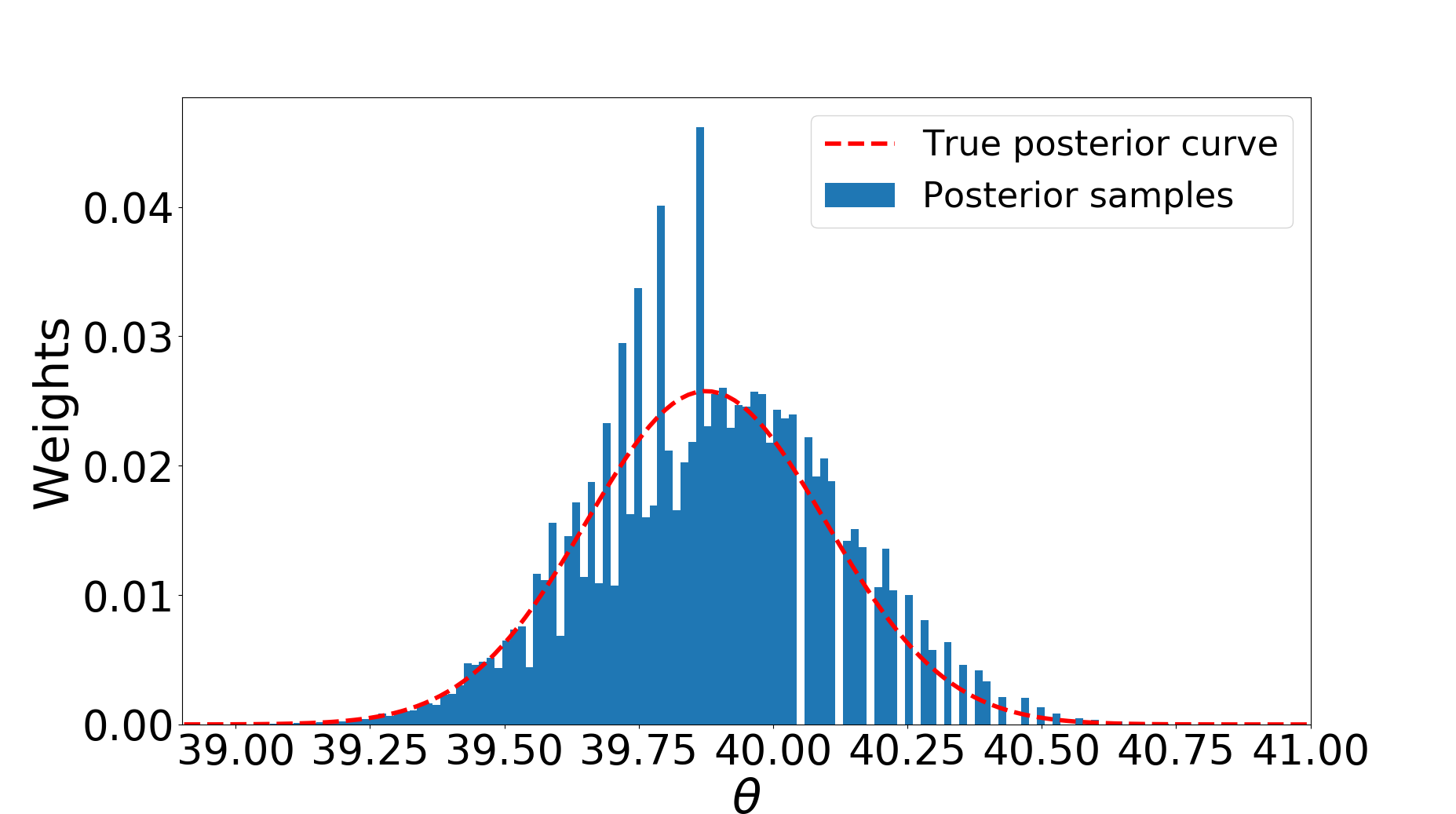}}
\hspace{0.5cm}\subfigure[$\mbeta = 0.4$]{
\includegraphics[width = 0.46\linewidth]{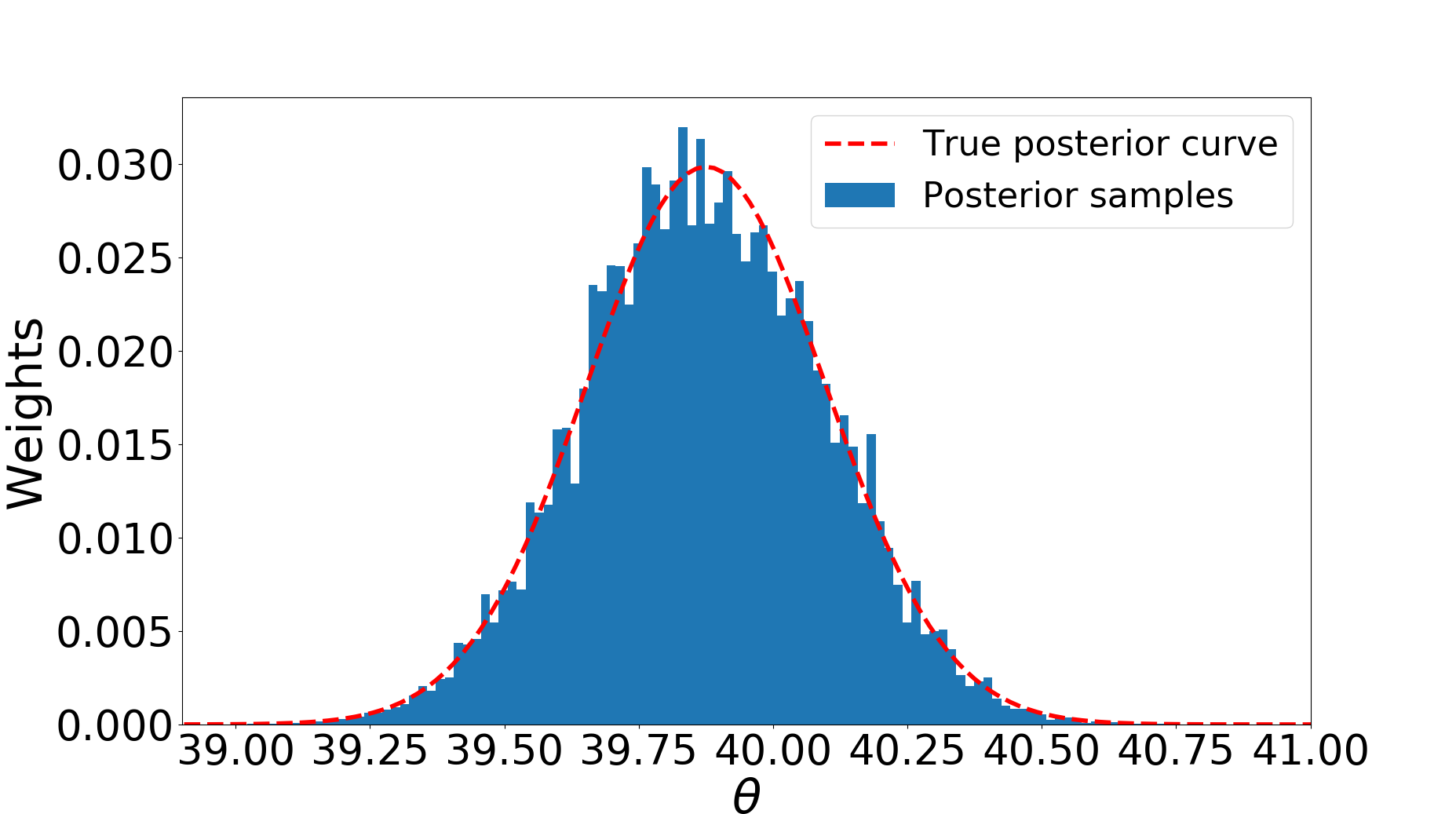}}
\subfigure[$\mbeta = 0.2$]{
\includegraphics[width = 0.46\linewidth]{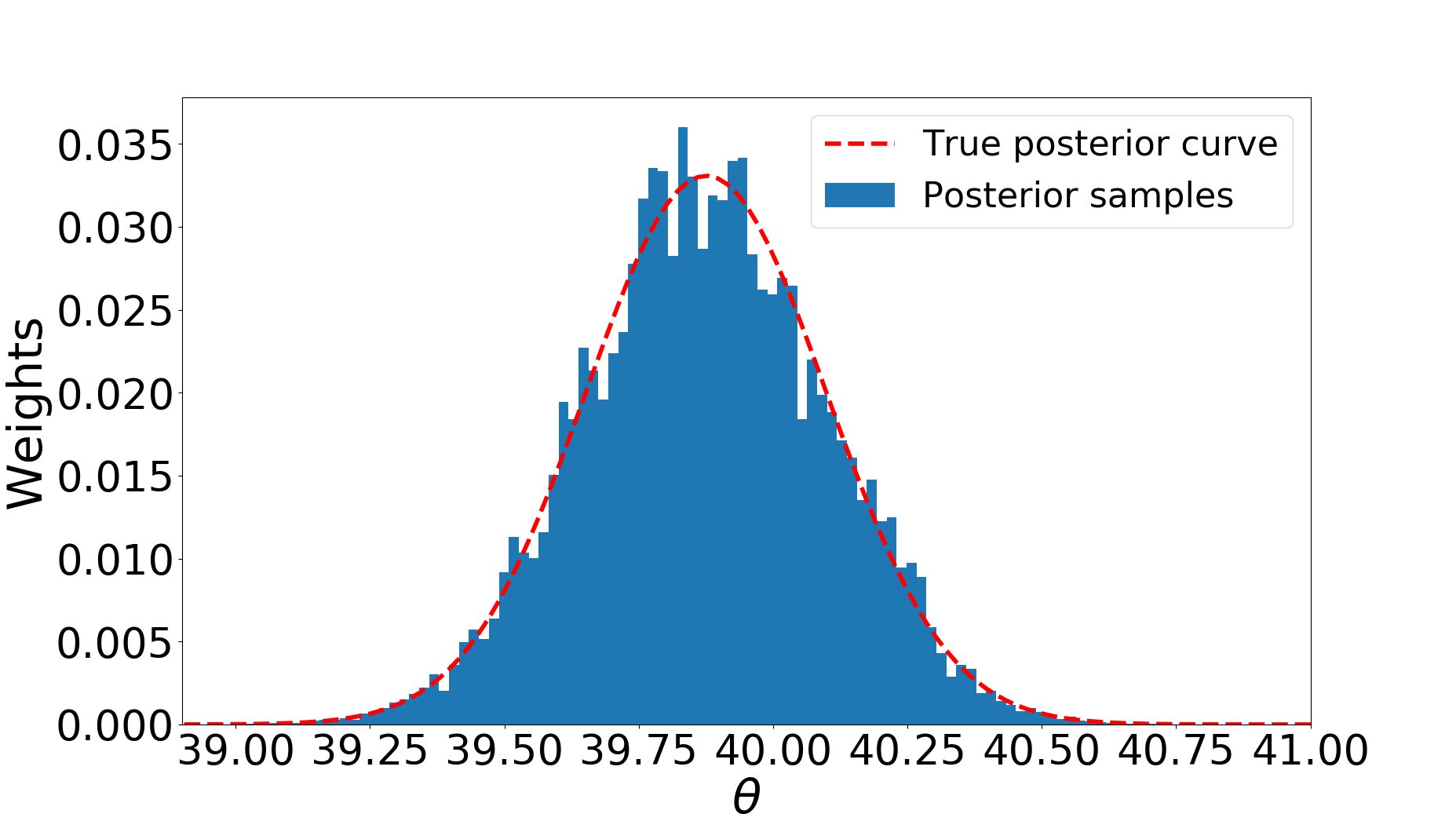}}
\hspace{0.5cm}\subfigure[$\mbeta = 0$, uniform prior]{
\includegraphics[width = 0.46\linewidth]{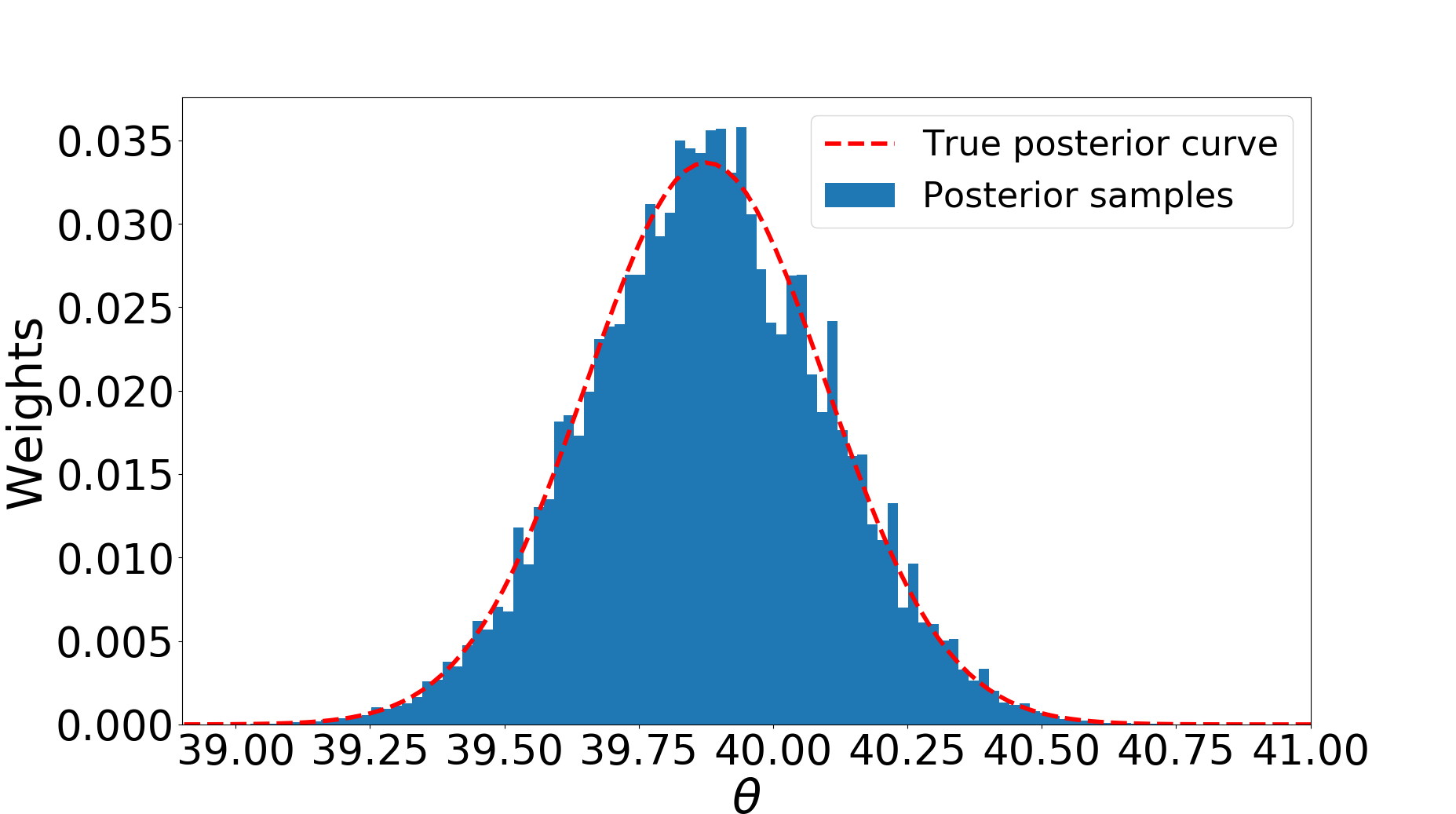}}
\caption{MultiNest performance using the PR method with different
  $\mbeta$ values, applied to case (3) ($\mtheta_\ast = 40$) of the
  toy example discussed in Section~\ref{Sec:simpExp}; all other
  settings remain unaltered. The values $\mbeta = 0.8, 0.6, 0.4, 0.2, 0$
  are tested. Figure (a) shows the distribution of the prior (blue
  dashed curve), likelihood (green dashed curve), ground truth (black
  dashed line), and posterior (red dashed curve). The remaining five
  figures show the histograms (blue bins)
of the posterior-weighted samples for the
  $\mbeta$ values tested and the true posterior distribution (red curve).}
\label{fig:exp1}
\end{figure*}

\revised{We use the open-source MultiNest package
  \citep{feroz2009multinest} and set efficiency parameter ${\rm efr} =
  0.8$, convergence tolerance parameter ${\rm tol} = 0.5$, multi-modal
  parameter ${\rm mmode} = {\rm False}$, random seed control parameter
  ${\rm seed}=-1$, and the constant efficiency mode ${\rm ceff} = {\rm
    False}$ for all the following examples. The number of live samples
  $N_{\rm live}$ varies in different cases. We keep the other
  MultiNest tuning options in their default values. See
  \citep{feroz2009multinest} and its corresponding MultiNest Fortran
  package for details of these default settings.} 

\revised{In some of the multi-dimensional cases, we also compare the MultiNest
performance with MCMC. Specifically, a standard Metropolis--Hastings
sampler is implemented and applied to the same numerical
examples. Other MCMC samplers such as No-U-Turn Sampler (NUTS), and
slice samplers give similar performance in the numerical examples. One
popular Python implementation of these samplers can be found in PyMC3
\citep{salvatier2016} package. In some cases, we also compare the
performance of importance sampling \citep{neal2001annealed,
  tokdar2010importance, martino2017group}, using a standard IS
implementation from Python package `pypmc' \citep{Jahn2018pypmc}.}

\subsection{Toy univariate example revisited}
Here we re-use case (3) of the toy example discussed in
Section~\ref{Sec:simpExp}, for which MultiNest was shown to fail
without applying PR. In this case, the true value of the unknown
parameter is $\mtheta_\ast = 40$ and the number of observations is set
to $N = 20$ (see Figure \ref{fig:exp1}(a)).

We use power prior redefinition and consider the $\mbeta$ values $0,
0.2, 0.4, 0.6, 0.8$ and $1$; note that $\mbeta = 1$ is equivalent to the
original method implemented in the toy example, and $\mbeta = 0$
corresponds to using a uniform distribution as the modified prior. The
range of the uniform prior for $\mbeta = 0$ is set as $\mtheta \in
[0,50]$ in this example.

Figure \ref{fig:exp1} shows the performance of MultiNest assisted by
the PR method. Panels (b) to (f) show the MultiNest performance with
decreasing $\mbeta$. One sees that as $\beta$ decreases, the posterior
samples obtained approximate the true posterior with increasing
accuracy, although in this extreme example one requires $\beta=0.4$ or
lower to obtain consistent results.

\begin{figure}[!ht]
\centering
\subfigure[case (1), $\mTheta_{\ast}=5$]{
\includegraphics[width = 0.9\linewidth]{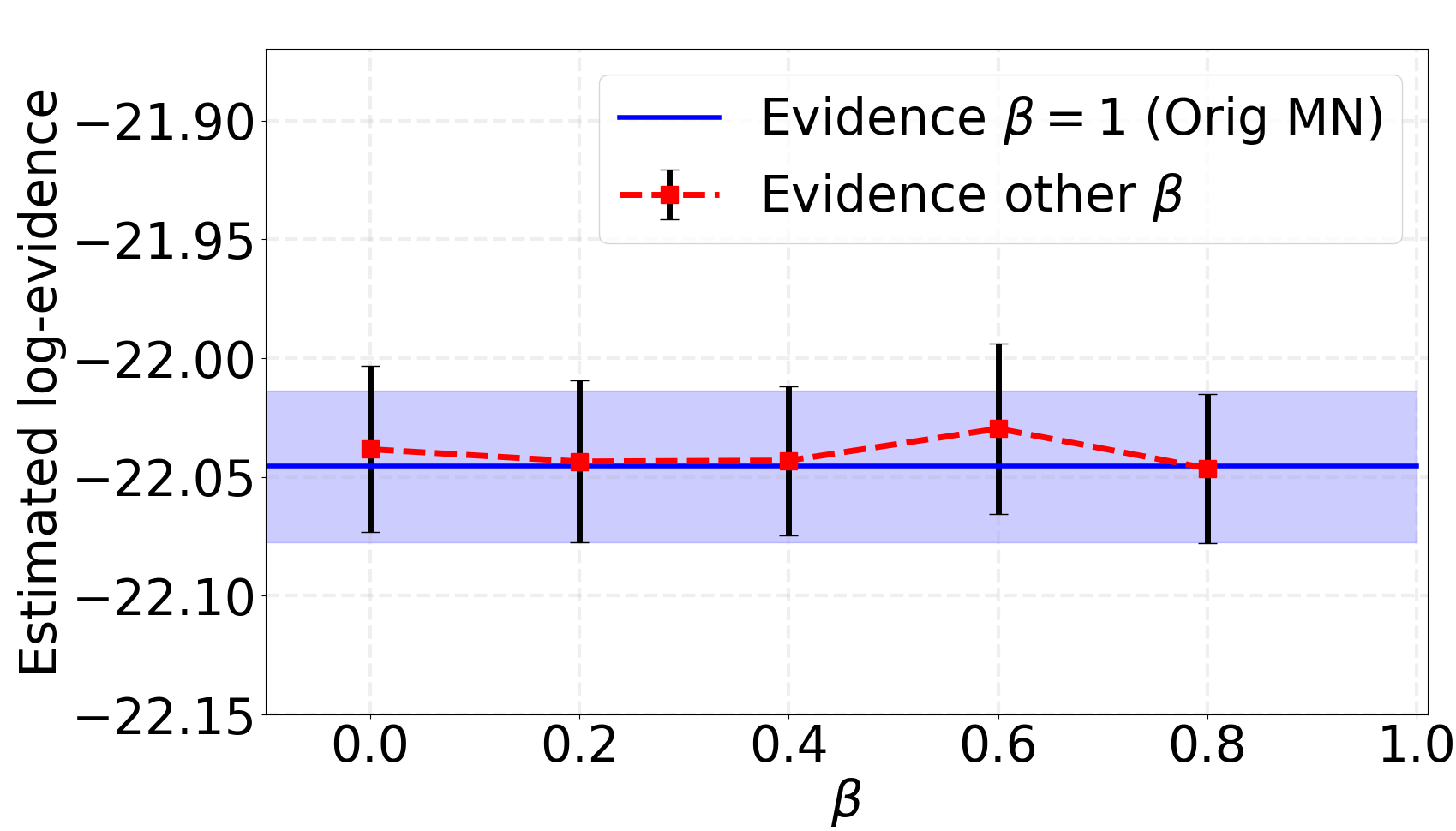}}
\subfigure[case (2), $\mTheta_{\ast}=30$]{
\includegraphics[width = 0.9\linewidth]{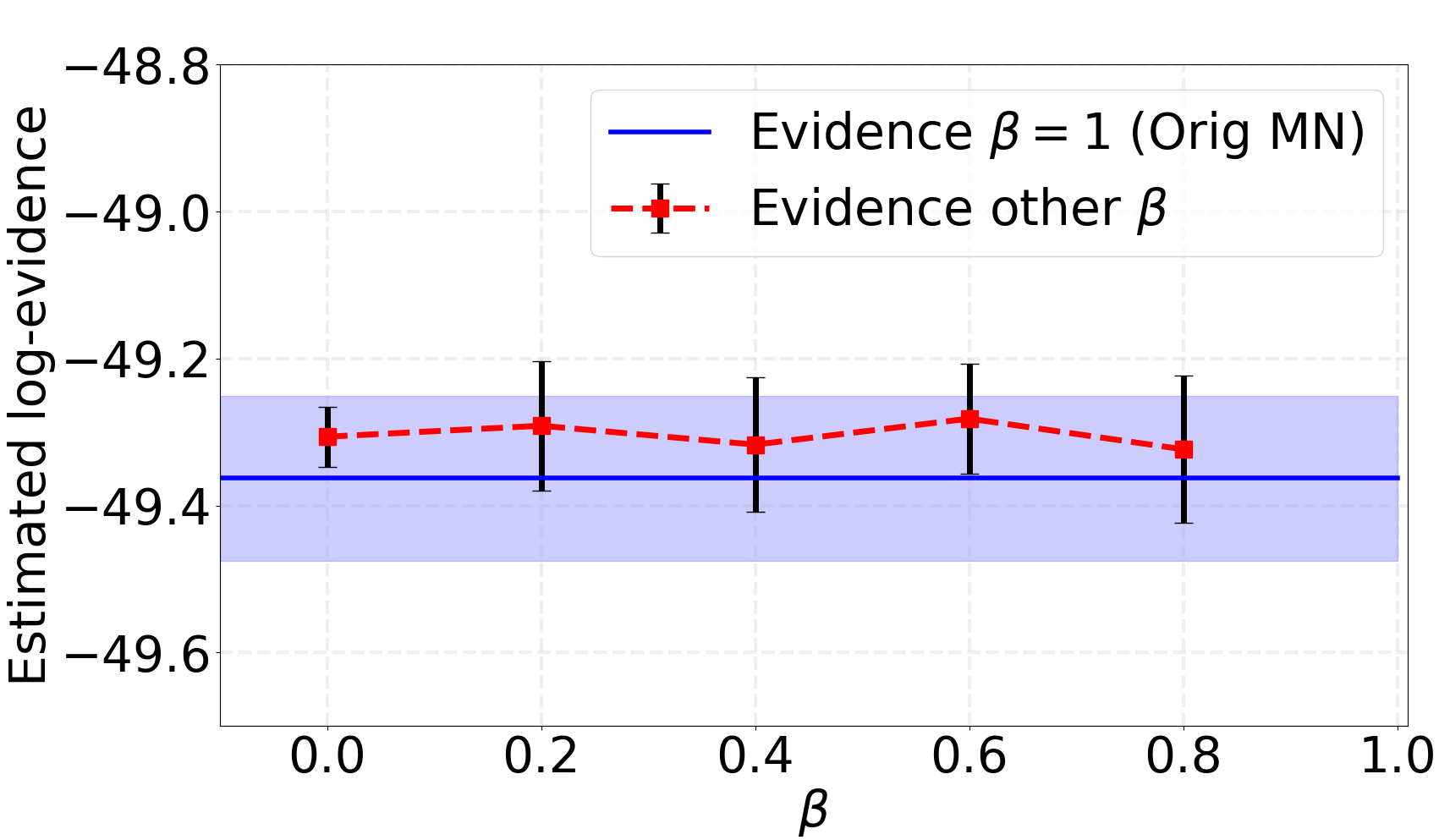}}
\subfigure[case (3), $\mTheta_{\ast}=40$]{
\includegraphics[width = 0.9\linewidth]{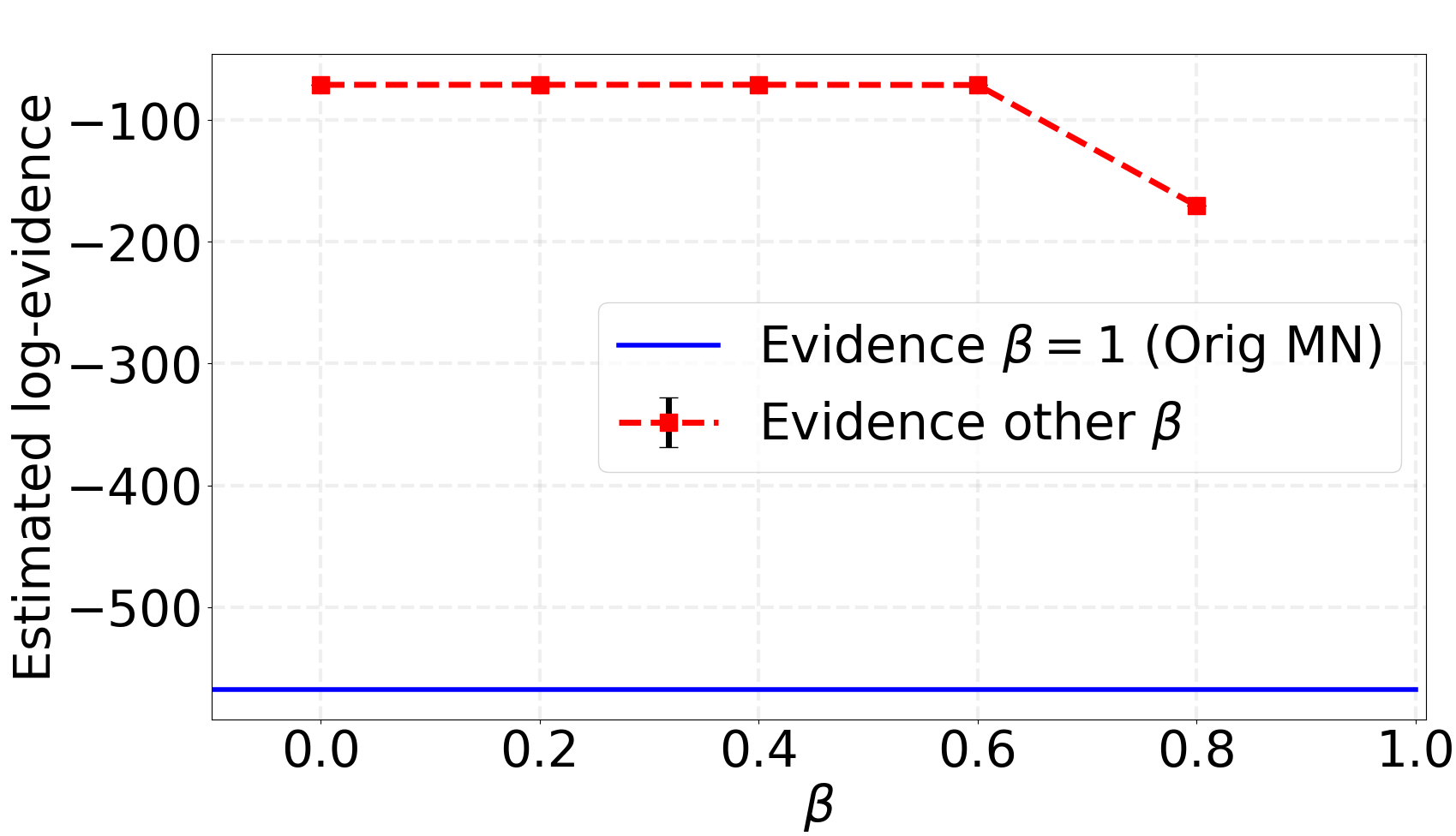}}
\caption{\revised{Evidence estimation versus $\mbeta$ for cases
    (1)--(3) of the univariate toy example. The blue solid line and
    the light blue shaded area indicate, respectively, the average and
    standard deviation of the log-evidence values produced by
    MultiNest without PR ($\mbeta = 1$) from 20 realisations of the
    data.  The red marker black cap errorbar shows the corresponding
    quantities produced using PR with $\mbeta = 0, 0.2, 0.4, 0.6, 0.8, 1$.}}
\label{fig:exp1Evi}
\end{figure}

To evaluate the performance of the PR method further, MultiNest was
run on 10 realisations for each value of $\mbeta$. The resulting
histograms of MultiNest's posterior samples were then fitted with a
standard Gaussian distribution. For each value of $\mbeta$, the
average of the means of the fitted Gaussian distributions and the root
mean squared error (RMSE) between these estimates and the true value
are presented in Table \ref{tab:exp1}, along with the average number
of likelihood calls for MultiNest to converge; since the time spent
for each likelihood calculation is similar, this quantity is
proportional to the runtime.  The RMSE clearly decreases as $\mbeta$
decreases from unity to zero, which demonstrates that a wider prior
allows MultiNest to obtain more accurate results, even in this extreme
example of an unrepresentative prior.  Also, one sees that the
averaged number of likelihood evaluations also decreases significantly
with $\mbeta$, so that the computational efficiency is also increased
as the effective prior widens. 

\begin{table}[h]
\centering
\caption{\revised{A numerical comparison of the results in the univariate toy
  example of the PR method for different values of $\mbeta$ (where
  $\mbeta=1$ corresponds to the standard method). The quantity
  $\bar{\mu}$ denotes the averaged mean value of the fitted Gaussian
  distribution to the posterior histogram over 10 realisations. RMSE
  denotes the root mean squared error between the ground truth value
  and $\bar{\mu}$. ${\rm N_{like}}$ is the averaged number of likelihood
  evaluations, and $\mZ$ denotes the averaged estimated log-evidence and its uncertainty given by MultiNest.}}
\begin{tabular}{c|c|c|c|c}
\hline
$\mbeta$ & $\bar{\mu}$ & RMSE & ${\rm N_{like}}$ & $\mZ$\\
\hline
1 & $32.838$ & $7.037$ & $96378$ & $-567.5679 \mypm 0.1346$\\
0.8 & $36.714$ & $3.161$ & $93492$ &$-170.3971 \mypm 0.1347$\\
0.6 & $39.870$ & $0.005$ & $83619$ & $-71.1709 \mypm 0.1276$\\
0.4 & $39.872$ & $0.003$ & $61796$ &$ -70.9523 \mypm 0.1269$\\
0.2 & $39.874$ & $0.001$ & $39013$ &$-70.9795 \mypm 0.0810$\\
0 & $39.875$ & $0.001$ & $15897$ & $-71.0134 \mypm 0.0441$\\
\hline
\end{tabular}
\label{tab:exp1}
\end{table}

These results illustrate the general procedure mentioned at the end of
Section~\ref{Sec:powerPR}, in which one obtains inferences for
progressively smaller values of $\mbeta$, according to some
pre-defined or dynamic `annealing schedule', until the results
converge to a statistically consistent solution. This is explored
further \revised{in the example considered in the next section}.


\revised{Before moving on, however, it is also of interest to
  investigate the evidence values obtained with and without the PR
  method. For completeness, we reconsider all three cases of the toy
  example discussed in Section~\ref{Sec:simpExp}, namely: (1)
  $\mTheta_{\ast} = 5$; (2) $\mTheta_{\ast} = 30$; and $\mtheta_\ast =
  40$. In each case, we calculate the mean and standard deviation of
  the log-evidence reported by MultiNest over 20 realisations of the
  data for $\mbeta = 0, 0.2, 0.4, 0.6, 0.8, 1$, respectively. The
  results are shown in Figure~\ref{fig:exp1Evi}, in which the blue
  solid line and the light blue shaded area indicate, respectively,
  the average and standard deviation of the log-evidence values
  produced by MultiNest without PR ($\mbeta = 1$), and the red marker
  black cap errorbar shows the corresponding quantities produced using
  PR with other $\mbeta$-values.}

\revised{For case (1), the red dashed curve fluctuates around the
  benchmark blue line at $-22.0182$ ($\mbeta = 1$ case), and the
  evidence estimates have similar size uncertainties, as one would
  expect. For case (2), however, one sees that the mean evidence
  values do change slightly as $\mbeta$ is decreased from unity,
  converging on a final value for $\mbeta < 0.8$ that is $\sim 0.1$
  log-units larger than its mean value for $\mbeta=1$. This indicates
  that case (2) also suffers (to a small extent) from the
  unrepresentative prior issue, despite this not being evident from
  the posterior samples plotted in Figure~\ref{fig:prior1}(d). For
  case (3), as expected, one sees that the mean log-evidence values
  change vastly as $\mbeta$ is decreased from unity, converging for
  $\mbeta < 0.6$ on a value that is $\sim 500$ log-units higher than
  for $\mbeta=1$. This large difference means that the error-bars are
  not visible in this case, so the mean and standard deviation of the
  log-evidence for each $\mbeta$ value are also reported in the last
  column of the Table~\ref{tab:exp1}.}

\revised{These results demonstrate that the PR method also works
  effectively for evidence approximation in nested sampling, as well
  as producing posterior samples. Indeed, it also suggests that the
  evidence might be a useful statistic to monitor for convergence as
  one gradually lowers the value of $\mbeta$ in the PR method.}

\subsection{Bivariate example}
\label{sec:bivariate}
As our second example we consider a bivariate generalisation of our
previous example, since it is straightforward to visualise. The
bivariate case can easily be extended to higher dimensionality. 

\begin{figure*}[!ht]
\centering
\subfigure[$\mTheta_\ast=(0.5,0.5)$]{
\includegraphics[width = 0.32\linewidth]{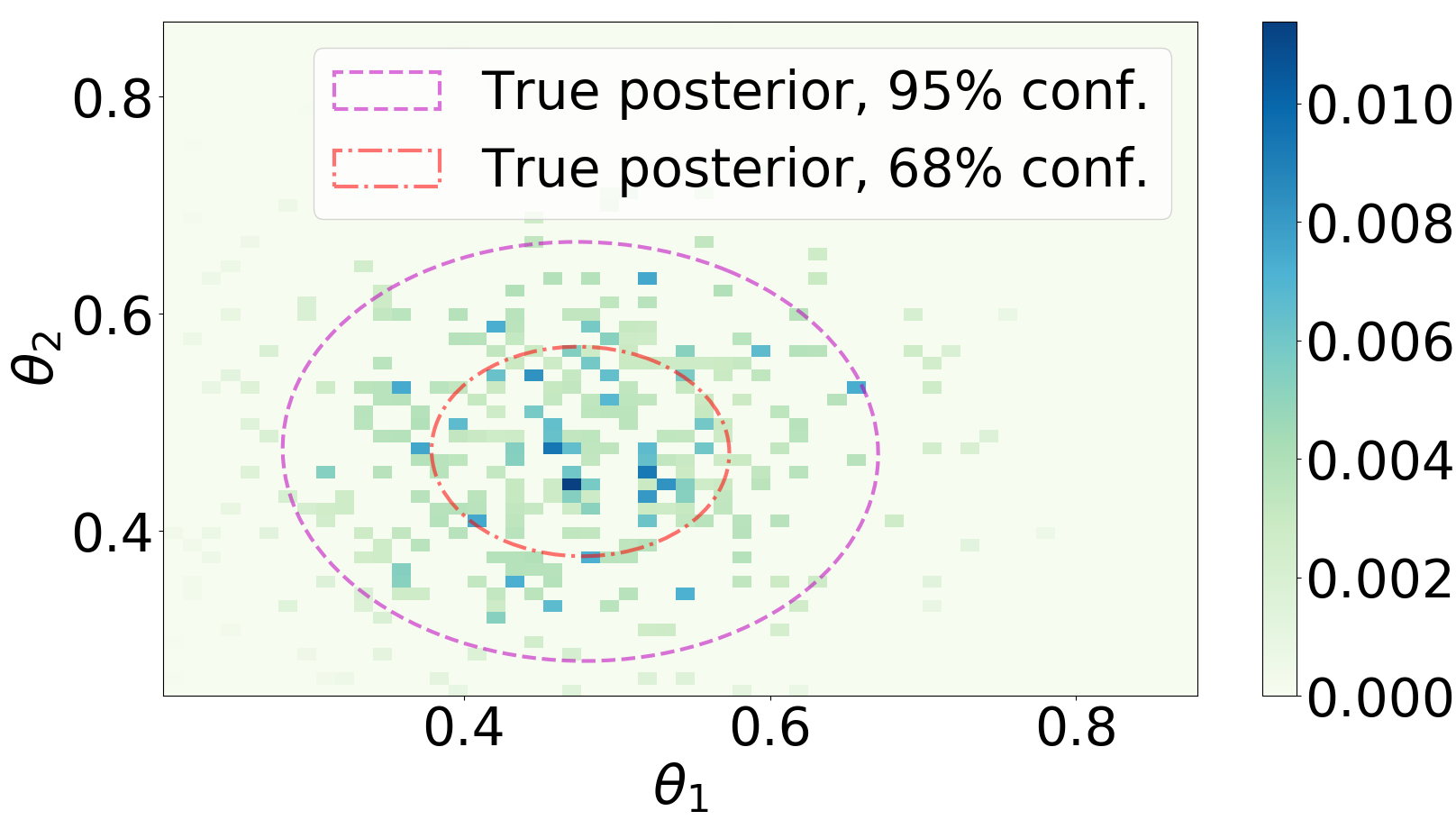}}
\subfigure[$\mTheta_\ast=(1.5,1.5)$]{
\includegraphics[width = 0.32\linewidth]{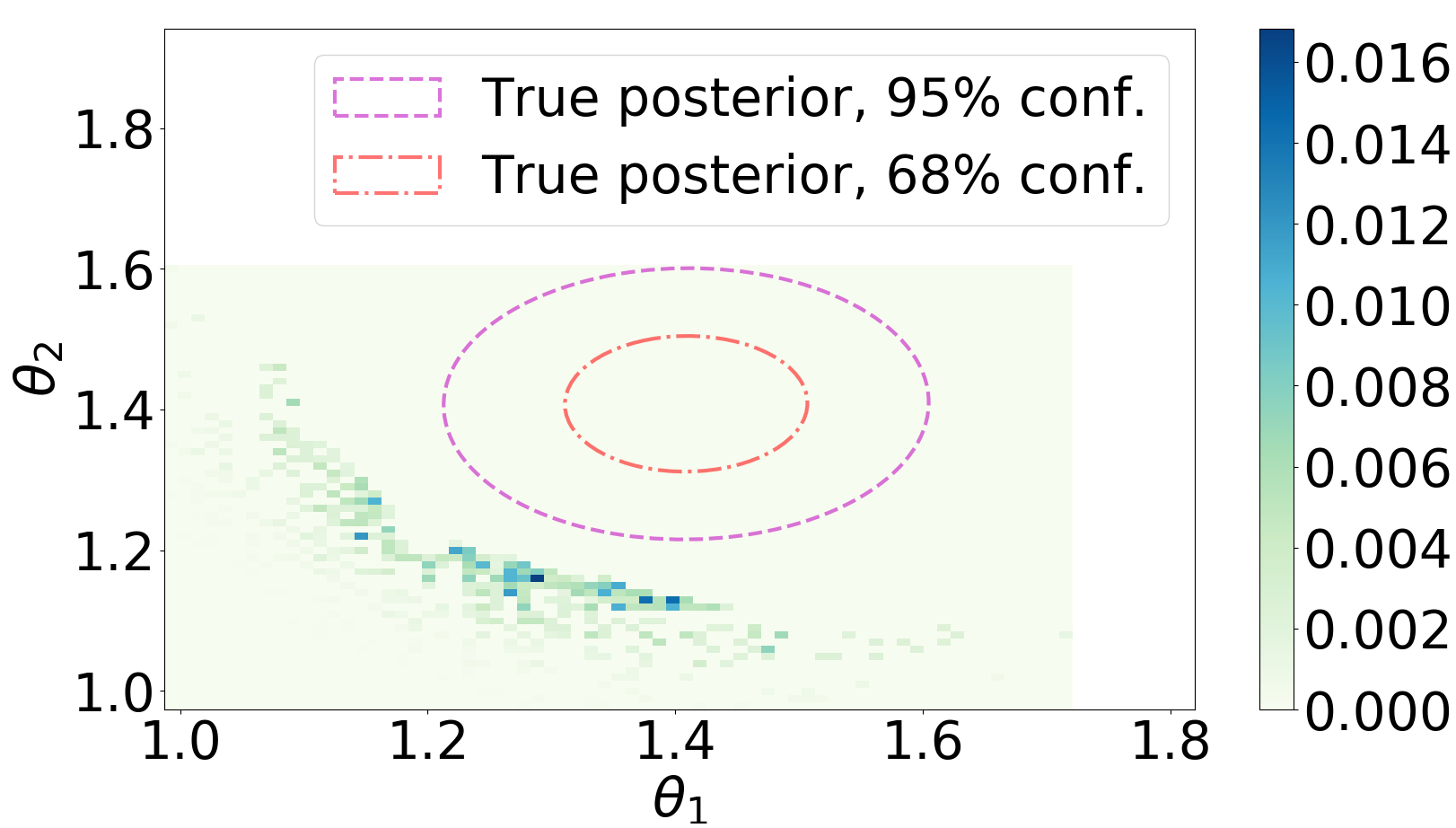}}
\subfigure[$\mTheta_\ast=(2.0,2.0)$]{
\includegraphics[width = 0.32\linewidth]{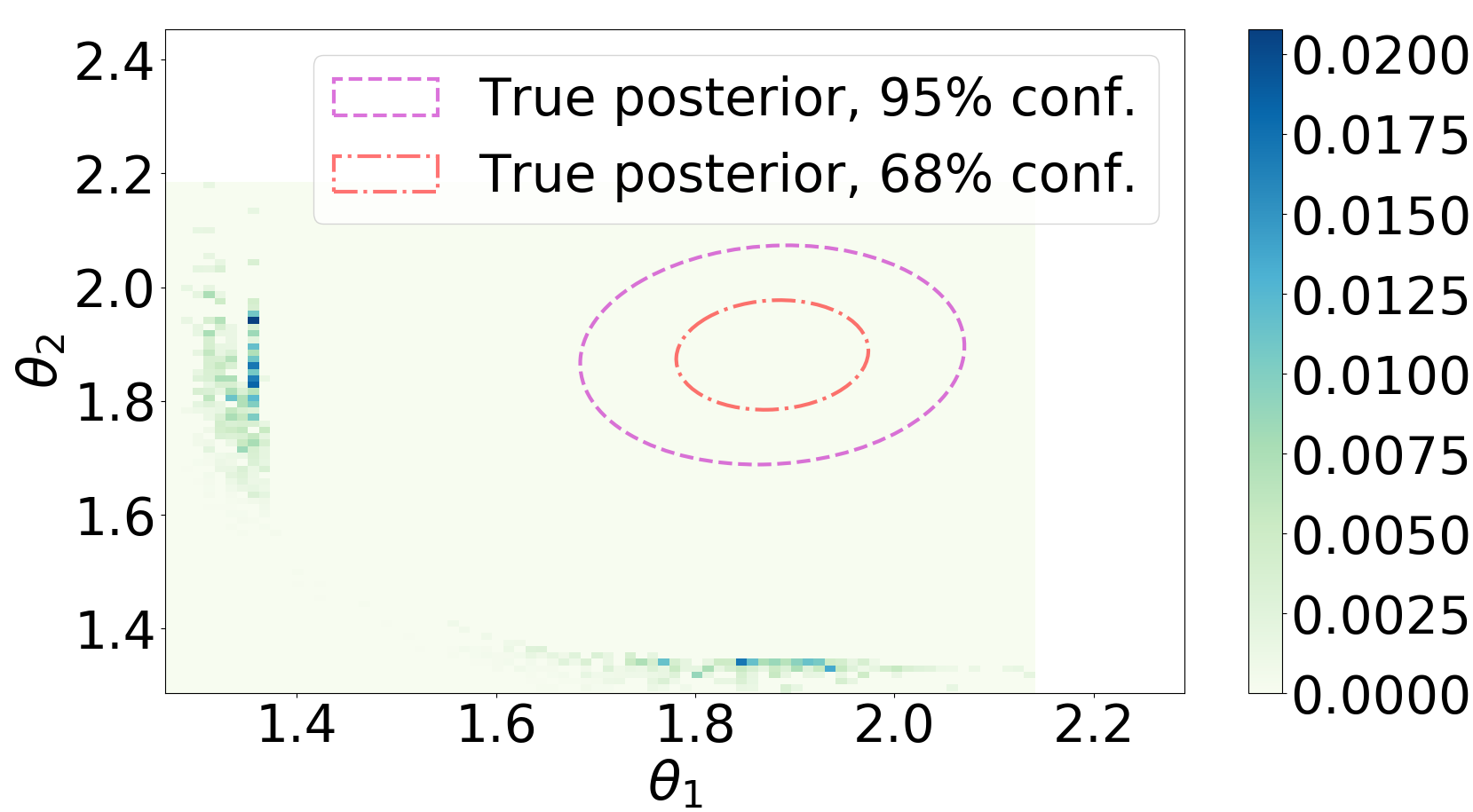}}
\caption{Two-dimensional histograms of MultiNest
    posterior samples (color scale) obtained without PR in the
    bivariate example, for cases (1)--(3). The colour map from light yellow to dark blue denotes low to high posterior sample density. The 68\% and 95\% contours
    of the true posterior distribution is each case are also shown.}
\label{fig:exp2overall}
\end{figure*}

Suppose one makes $N$ independent measurements $\mathbf{X} =
[\mathbf{x}_{1}, \cdots, \mathbf{x}_{n}, \cdots, \mathbf{x}_{N}]^{\top}$ of
some two-dimensional quantity $\mTheta=(\mtheta_1,\mtheta_2)^{\top}$, such
that in an analogous manner to that considered in equation
\eqref{Eq:Simpexp} one has
\begin{align}
\mathbf{x}_{n} = \mTheta + \mXi, \label{Eq:Biexp}
\end{align}
where $\mXi=(\mxi_1,\mxi_2)$ denotes the simulated measurement noise,
which is Gaussian distributed $\mXi \sim \calN (\mmu_{\mXi},
\mSigma_{\mXi})$ with mean $\mmu_{\mXi}$ and covariance matrix
$\mSigma_{\mXi}$.  For simplicity, we will again assume the
measurement process is unbiased, so that $\mmu_{\mXi} = (0,0)$, and
that the covariance matrix is diagonal
$\mSigma_{\mXi}=\mbox{diag}(\sigma_{\mxi_1}^2,\sigma_{\mxi_2}^2)$, so
that there is no correlation between $\mxi_1$ and $\mxi_2$, and the
individual variances are known {\em a priori}. We also assume a
bivariate Gaussian form for the prior \revised{$\mTheta \sim \calN
(\mmu_{\mTheta}, \mSigma_{\mTheta})$}, where $\mmu_{\mTheta} = (0,0)$
and
$\mSigma_{\mTheta}=\mbox{diag}(\sigma_{\mtheta_1}^2,\sigma_{\mtheta_2}^2)$.

We consider three cases, where the true values of the unknown
parameters are, respectively, given by: (1) $\mTheta_\ast=(0.5,0.5)$;
(2) $\mTheta_\ast=(1.5,1.5)$; and (3) $\mTheta_\ast=(2.0,2.0)$. In
each case, we assume the noise standard deviation to be
$\sigma_{\mxi_1}=\sigma_{\mxi_2}=0.1$, and the width of the prior to
be $\sigma_{\mtheta_1}=\sigma_{\mtheta_2}=0.4$. We assume one
observation for each case, i.e., $N=1$. 

In each case, the MultiNest sampling parameters were set to $N_{\rm
  live} = 100$, ${\rm efr} = 0.8$ and ${\rm tol} = 0.5$ (see
\citealt{feroz2009multinest} for details), and the algorithm was run
to convergence. The results obtained without apply the PR method
(which is equivalent to setting $\mbeta=1$) are shown in Figure
\ref{fig:exp2overall}. One sees that the MultiNest samples are
consistent with the true posterior distribution for case (1), but the
sampler fails in cases (2) and (3) in which the ground truth lies far
into the wings of the prior.

\begin{figure*}[ht]
\centering
\subfigure[case (2), $\mbeta = 0.6$]{
\includegraphics[width = 0.32\linewidth]{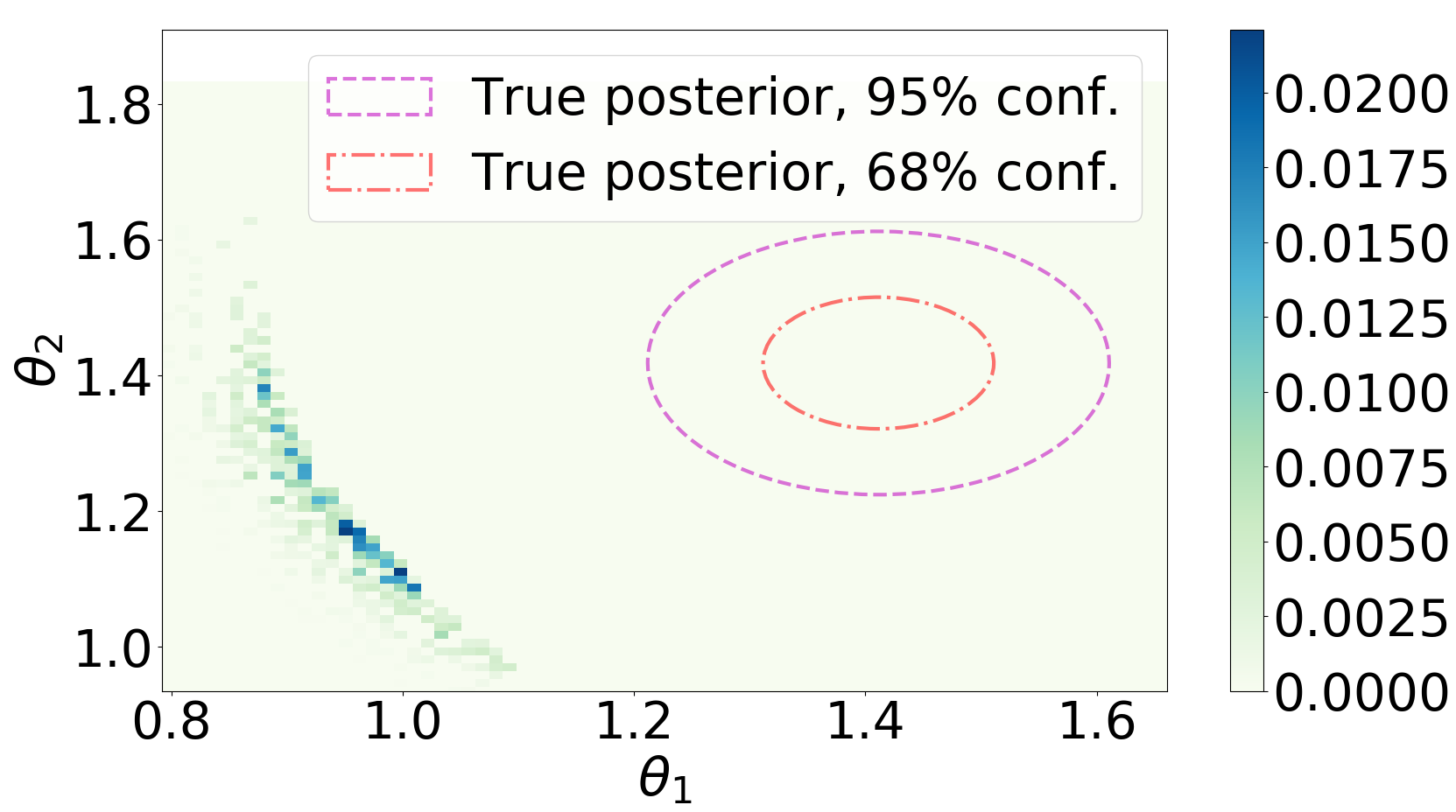}}
\subfigure[case (2), $\mbeta = 0.3$]{
\includegraphics[width = 0.32\linewidth]{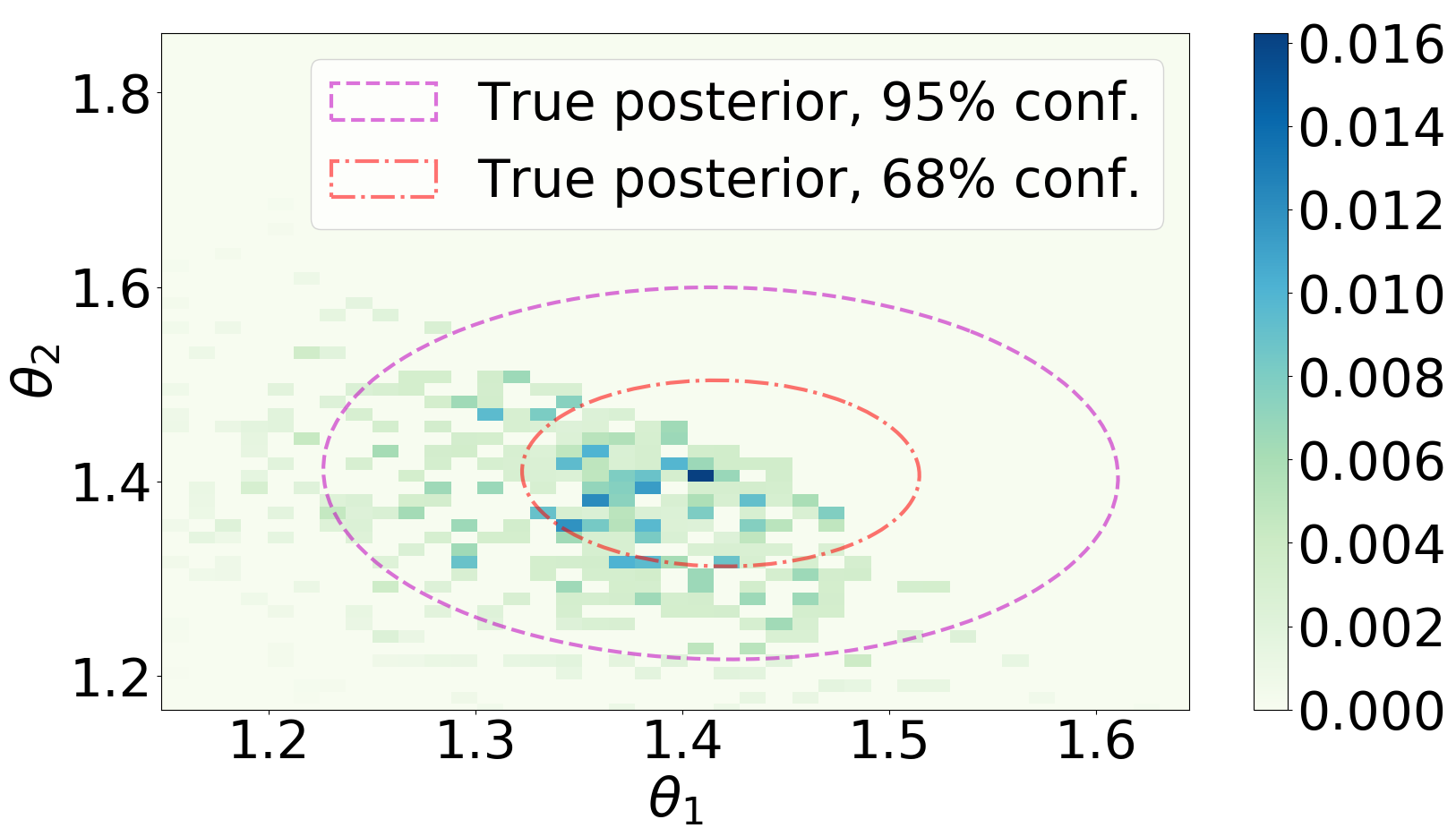}}
\subfigure[case (2), $\mbeta = 0.1$]{
\includegraphics[width = 0.32\linewidth]{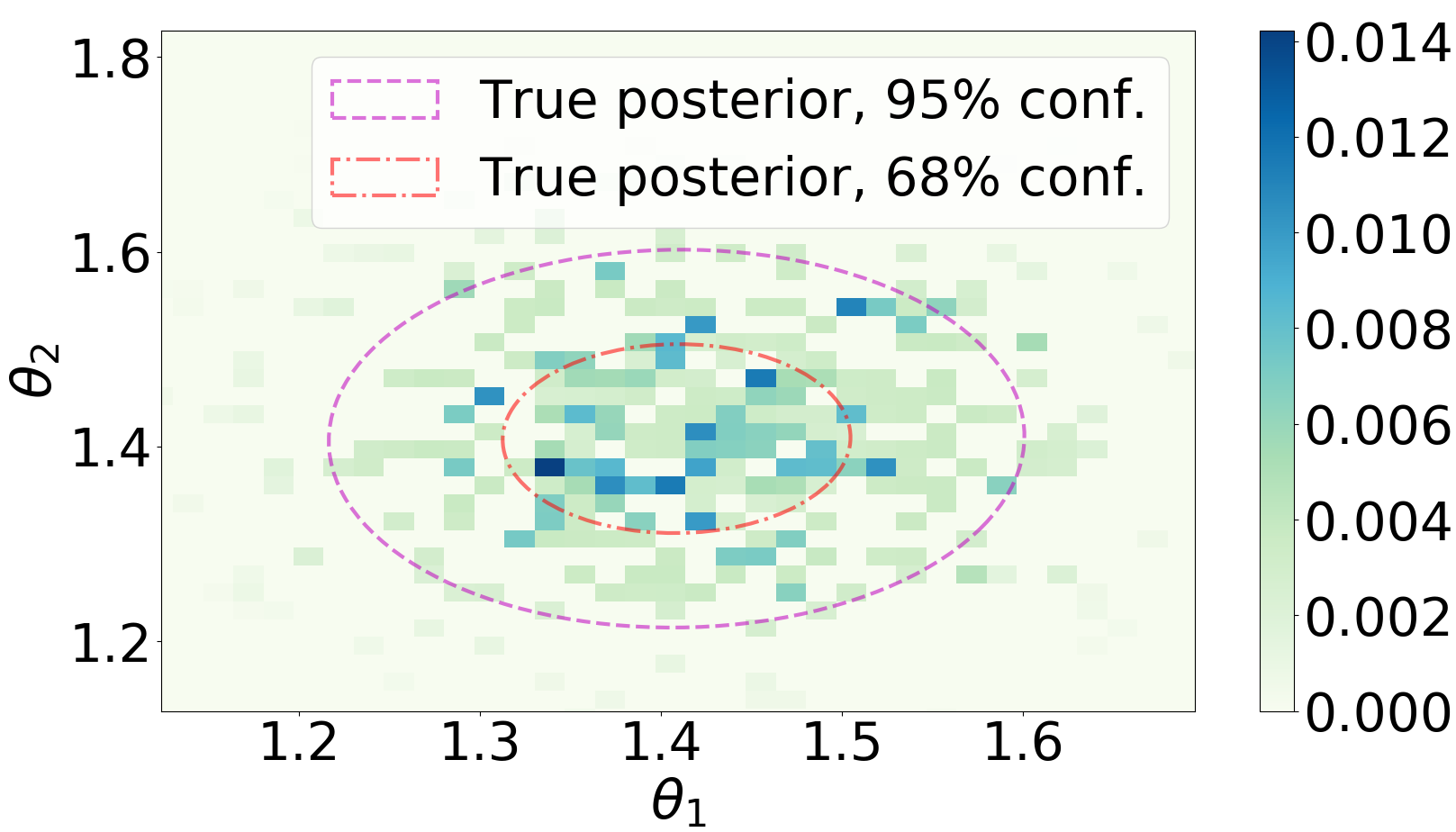}}
\subfigure[case (3), $\mbeta = 0.6$]{
\includegraphics[width = 0.32\linewidth]{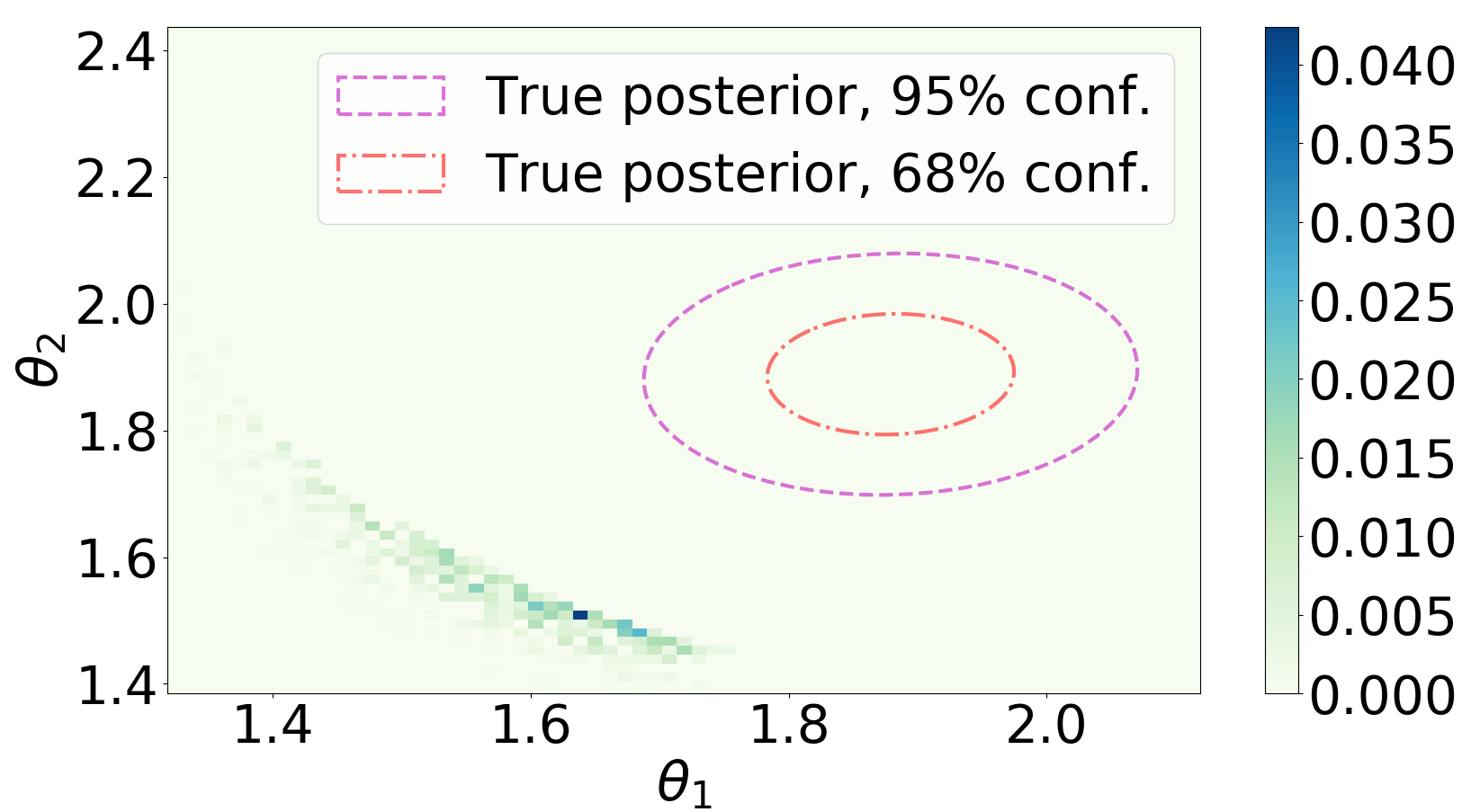}}
\subfigure[case (3), $\mbeta = 0.3$]{
\includegraphics[width = 0.32\linewidth]{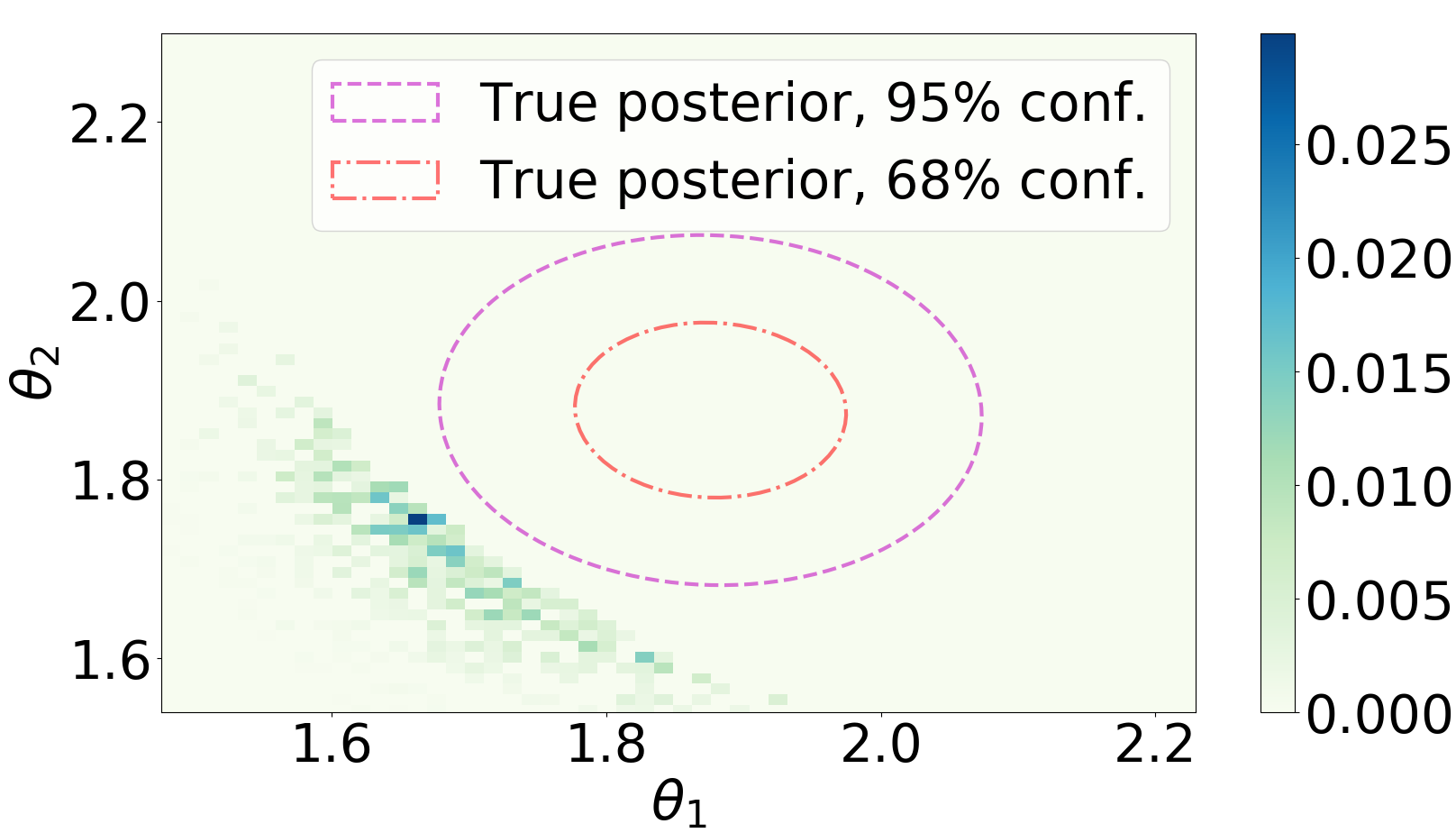}}
\subfigure[case (3), $\mbeta = 0.1$]{
\includegraphics[width = 0.32\linewidth]{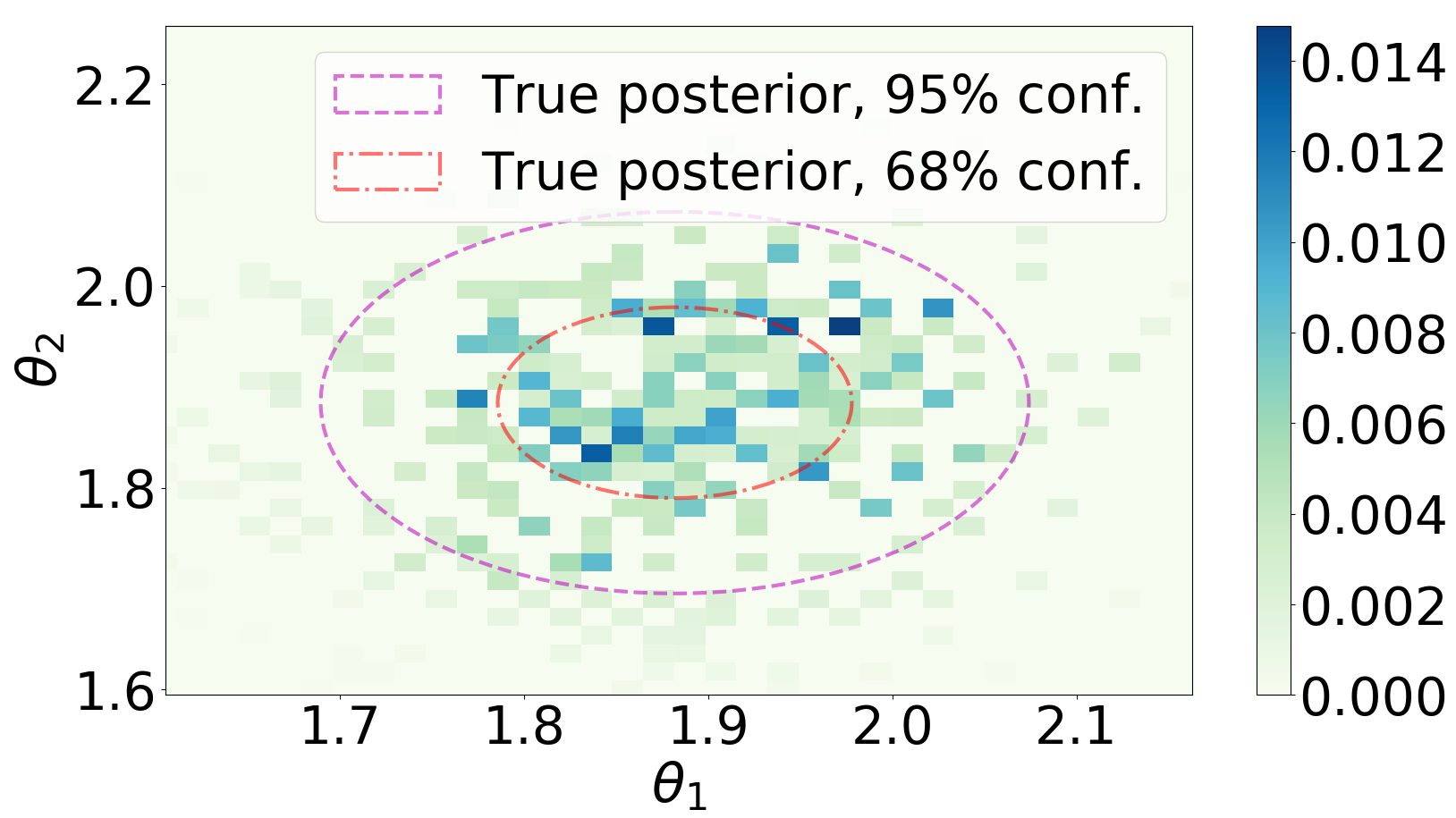}}
\caption{MultiNest performance with PR method in the bivariate toy
  example for case (2) $\mTheta_\ast=(1.5,1.5)$ (top four sub-figures)
  and case (3) $\mTheta_\ast=(2.0,2.0)$ (bottom four sub-figures). As
  indicated, the panels correspond to $\mbeta$ values of 0.6, 0.3, and 0.1, respectively. The colour map from light yellow to dark blue denotes low to high posterior sample density.}
\label{fig:exp2for1dot5}
\end{figure*}

\begin{table*}[ht]
\centering
\caption{\revised{A comparison between MultiNest with and without PR
    method (for various values of $\mbeta$, and 100 live samples),
    standard MCMC algorithm (termed as `MCMC'), and standard
    importance sampling algorithm (termed as 'IS') in the bivariate
    toy example for all three cases. The top half of the table is a
    comparison of RMSE, and the second half is for the number of
    likelihood evaluations ($N_{\rm like}$) per individual algorithm
    run. For $\mbeta = 0.05$ (highlighted in bold) MultiNest achieves
    relatively better and consistent RMSE and $N_{\rm like}$
    performance across the three cases. The number of posterior
    samples from the competing algorithms (MCMC and IS) are fixed to
    values around $1100$ in order to obtain a similar number of
    likelihood evaluations as required by MultiNest for $\mbeta =
    0.05$.}}
\begin{tabular}{c|cccccc|c|c}
\hline
RMSE & MN ($\mbeta=1$)  & $\mbeta=0.4$ & $\mbeta=0.2$ &
$\mbeta=0.1$ & $\mbeta=0.05$ & $\mbeta=10^{-5}$   &  MCMC & IS \\
\hline

Case (1) &  0.0066   &  0.0046 &  0.0055 &  0.0043  & \textbf{0.0038} &  0.0037   &  0.0293 &0.0252 \\
Case (2) &  0.3495   &  0.0518 &  0.0117&  0.0052  &  \textbf{0.0049} &  0.0046  &  0.0797 &  0.4117\\
Case (3) &  0.5586  &  0.3785 &  0.0276 &  0.0055  &  \textbf{0.0045} &  0.0044 &  0.0992 &0.8386 \\
\hline

$N_{\rm like}$  \\
\hline
Case (1) &  908   &  847 &  909 &  959  &  \textbf{1052}  &  2246  & 1100 &  1100\\
Case (2) &  2232  &  1553 &  1221 &  1127  &  \textbf{1118} &  2271  & 1100 &  1100\\
Case (3) &  3466  &  1922 &  1516 &  1280  &  \textbf{1188} &  2348  & 1100 &  1100\\
\hline
\end{tabular}
\label{tab:exp2}
\end{table*}

The MultiNest posterior samples obtained using the PR method, with
$\mbeta = 0.6, 0.3, 0.1$, respectively, are shown in
Figure \ref{fig:exp2for1dot5} for case (2) and case (3). In each case,
one sees that as $\mbeta$ decreases the samples become consistent with
the true posterior. In practice, it is thus necessary to reduce the
value of $\mbeta$ until the inferences converge to a sufficient
accuracy.

Table \ref{tab:exp2} summarises the inference accuracy and the
computational efficiency for all three cases for MultiNest without PR
(which corresponds to $\beta=1$) and with PR for $\mbeta = 0.4, 0.2,
0.1, 0.05, 10^{-5}$. One sees that for case (1)
$\mTheta_\ast=(0.5,0.5)$, applying PR to MultiNest has only a weak
effect on the RMSE performance and the number of likelihood
evaluations, with both changing by about a factor of about two (in
opposite directions) across the range of $\mbeta$ values considered.
For case (2) $\mTheta_\ast=(1.5,1.5)$ and case (3)
$\mTheta_\ast=(2,2)$, however, MultiNest without PR suffers from the
unrepresentative prior problem and the corresponding RMSE and number
of likelihood evaluations are considerably higher than in case
(1). Nonetheless, by combining MultiNest with the PR method, the RMSE
and number of likelihood evaluations can be made consistent across the
three cases considered.  One sees that the RMSE decreases as $\mbeta$
decreases and the maximum accuracy is obtained when $\mbeta = 10^{-5}$
(for which the modified prior is very close to uniform).  This should
be contrasted with the total number of likelihood evaluations, which
increases as $\beta$ decreases. Indeed, it is clear that the minimum
number of likelihood evaluations are required for intermediate values
of $\mbeta$. These results show that a reasonable compromise between
accuracy and computational efficiency is obtained for $\mbeta=0.05$ in
this problem, which also provides the best consistency for both RMSE and
the number of likelihood evaluations across all three cases.

\begin{figure*}[ht]
\centering
\subfigure[case (1), KL score 169]{
\includegraphics[width = 0.48\linewidth]{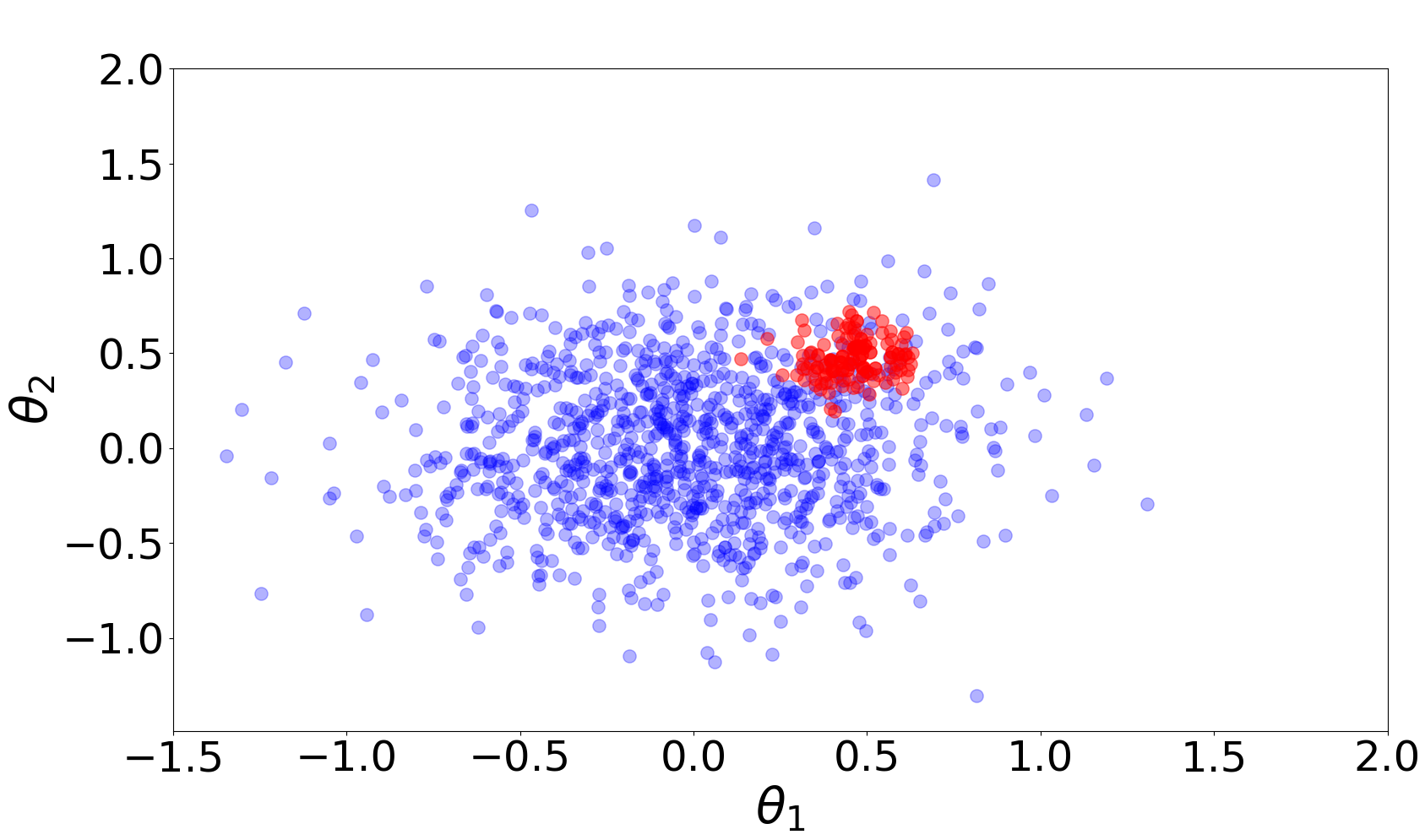}}
\subfigure[case (2), KL score 1833]{
\includegraphics[width = 0.48\linewidth]{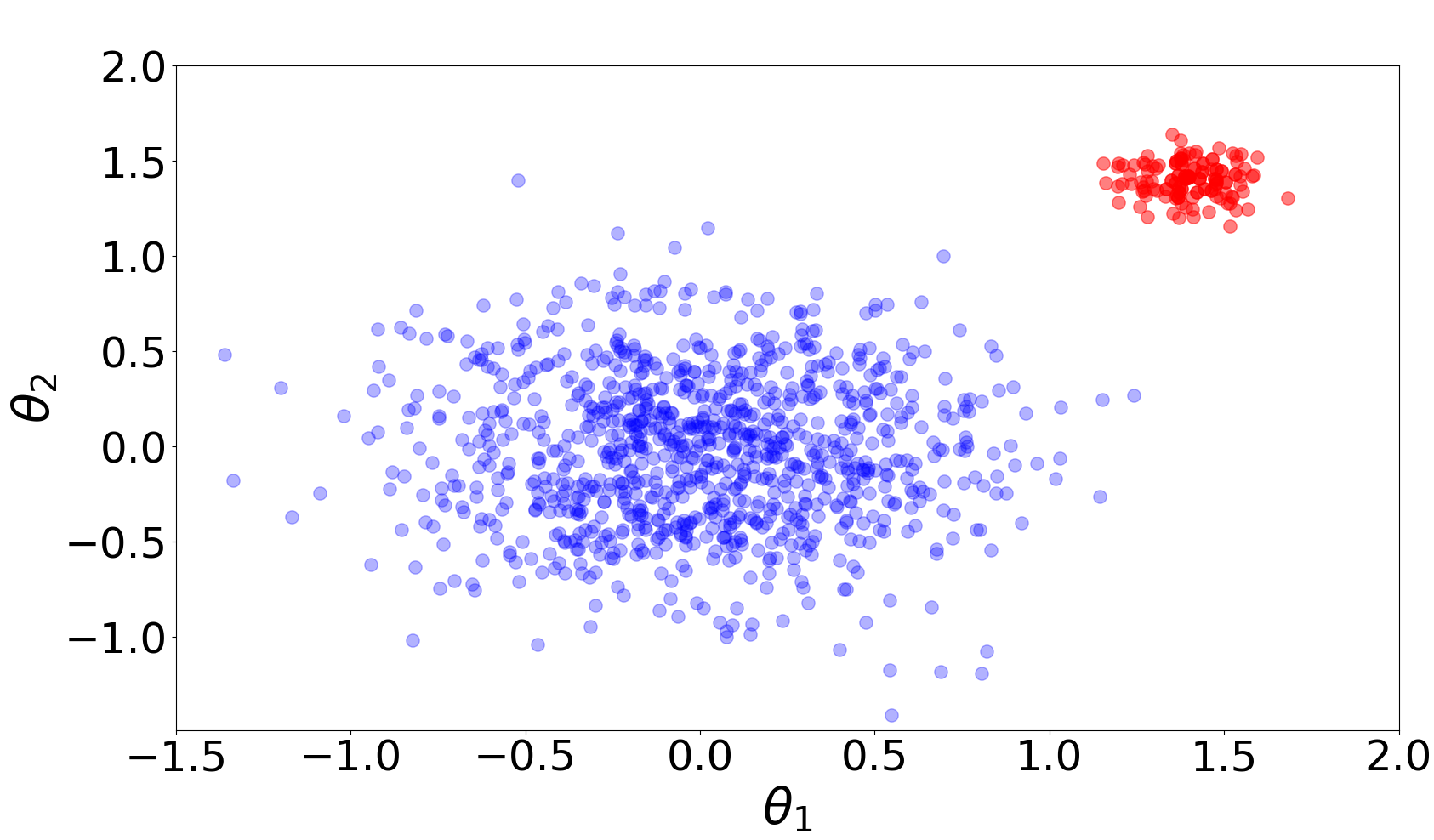}}
\caption{Demonstration of the KL divergence diagnostic for case (1)
  and case (2) in the bivariate example. The blue dots represent
  random samples drawn from the prior distribution, and the red dots
  are posterior samples from MultiNest with
  $\beta=0.01$ and $ N_{\rm live} = 100$. }
\label{fig:kl}
\end{figure*}

\subsubsection{Other sampling algorithms}

\revised{Since our focus here is to introduce the PR method to improve
  NS performance in problems with unrepresentative priors, a full
  comparison between MultiNest and other competing sampling algorithms
  is beyond the scope of this paper. Nonetheless, we report here on
  some results of a brief such comparison on the same bivariate
  example. In particular, we perform a comparison of MultiNest with
  PR, in terms both of the RMSE and the number of likelihood
  evaluations, with our own implementation of standard MCMC sampling
  using the Metropolis--Hastings algorithm with a Gaussian proposal
  distribution and also with importance sampling (IS), using a standard
  IS implementation from Python package `pypmc'
  \citep{Jahn2018pypmc}.}

\revised{The results obtained using MCMC and IS are shown in the final
  two columns of Table \ref{tab:exp2}. For $\mbeta = 0.05$ MultiNest
  achieves relatively better and consistent RMSE and ${\rm N_{like}}$
  performance across the three cases. The number of posterior samples
  from the competing algorithms (MCMC and IS) are fixed to values
  around $1100$ in order to obtain a similar number of likelihood
  evaluations as required by MultiNest for $\mbeta = 0.05$. For case
  (1), the performance of IS is comparable to that of MCMC. However,
  in cases (2) and (3), IS is comparable to MultiNest with $\mbeta =
  1$, so it is clear that IS also suffers from the unrepresentative
  prior problem.}

\revised{The detailed comparison of different sampling algorithms is a
  broad topic that has been widely discussed and explored in the
  literature. For example, importance sampling was formulated as a
  special case of bridge sampling, and was compared in
  \citep{gronau2017tutorial}. An importance nested sampling was
  proposed to incorporate importance sampling into NS evidence
  calculation step in \citep{feroz2013importance}. A comparison
  between NS and MCMC was discussed in
  \citep{allison2013comparison}. A review of importance sampling is
  presented in \citep{tokdar2010importance}.}

\subsubsection{Diagnostics for bivariate example}

We take the opportunity here to illustrate the diagnostics process
discussed in Section~\ref{sec:diagnostics} using the bivariate
example. Since case (1) does not suffer from the unrepresentative
prior problem, it can be treated as a reliable example and we assume
that the `available knowledge' is gained by analysing this case.  As
shown in Table \ref{tab:exp2}, the number of likelihood evaluations
(which is proportional to the runtime) for MultiNest without PR
$(\mbeta=1)$ increases significantly from case (1) to case (3). Thus,
the unrepresentative prior problem can be identified on-the-fly by
monitoring the runtime. An on-the-fly convergence rate check may also
be straightforwardly applied using existing rate of convergence
methods \citep{suli2003} to the problem. In either case, one may
identify that case (2) and case (3) differ significantly from the
available knowledge, and hence the PR method should be applied.

However, for some sampling methods, on-the-fly
diagnostic of monitoring the runtime would fail in the case (adopted
here) in which the number of likelihood evaluations is fixed. In this
case, one must therefore rely on an after-run diagnostic, such as the
KL divergence, which quantifies the differences between the assumed
prior and the corresponding posterior obtained in the analysis.
Figure \ref{fig:kl} shows MultiNest samples from the prior and the
posterior for case (1) and case (2), respectively, of the bivariate
example. By computing the standard KL divergence, we find a value
(termed KL score) of 169 for case (1) (the available knowledge) and
1833 for case (2). It is clear that the KL score for the
unrepresentative prior problem is much larger than normal case, and
so case (2) could be flagged as an outlier according to some
predefined criterion on KL score. Similarly considerations apply to
case (3).

\begin{figure*}[ht]
\centering
\subfigure[RMSE]{
\includegraphics[width = 0.48\linewidth]{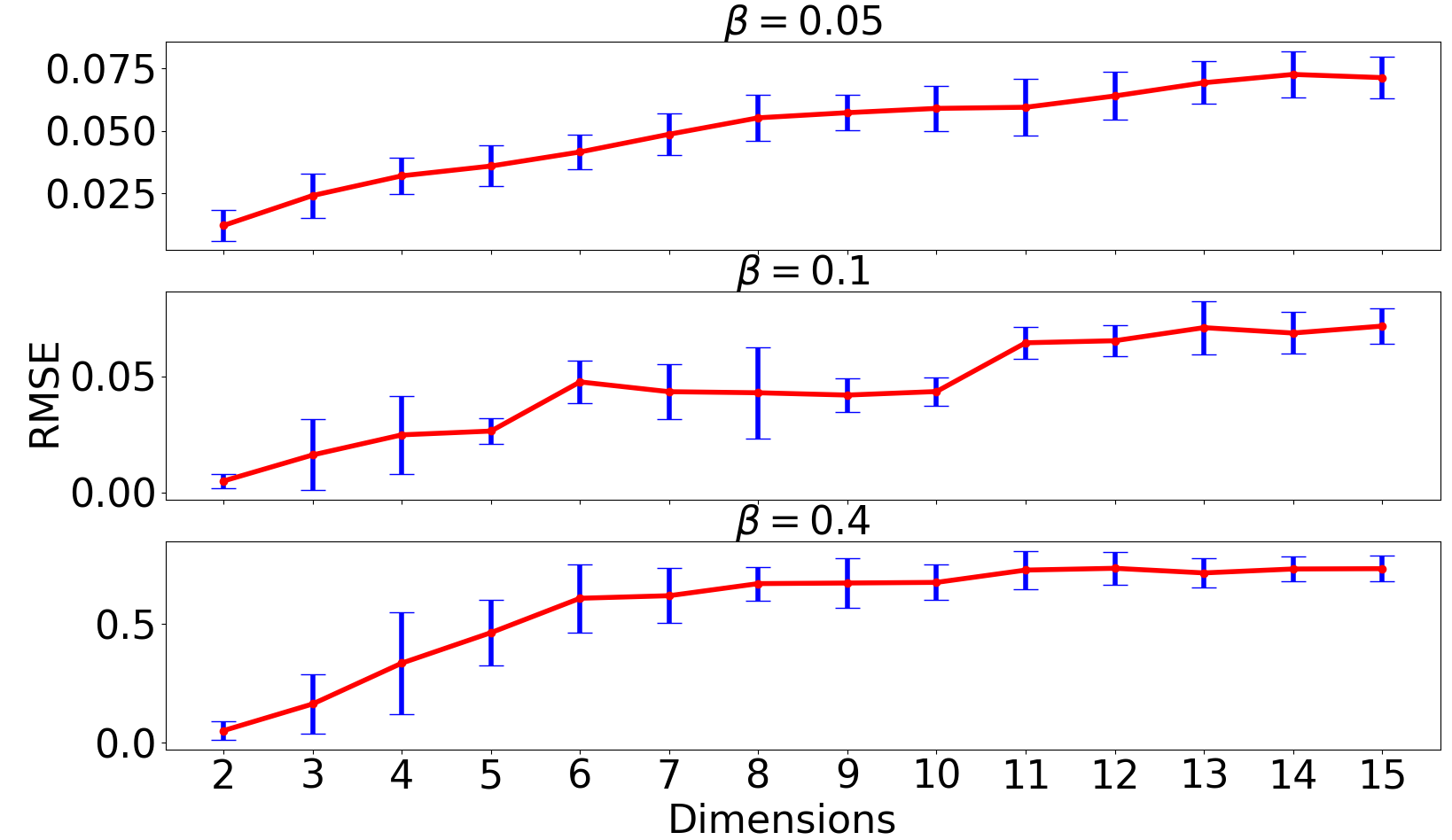}}
\subfigure[Number of likelihood evaluations]{
\includegraphics[width = 0.48\linewidth]{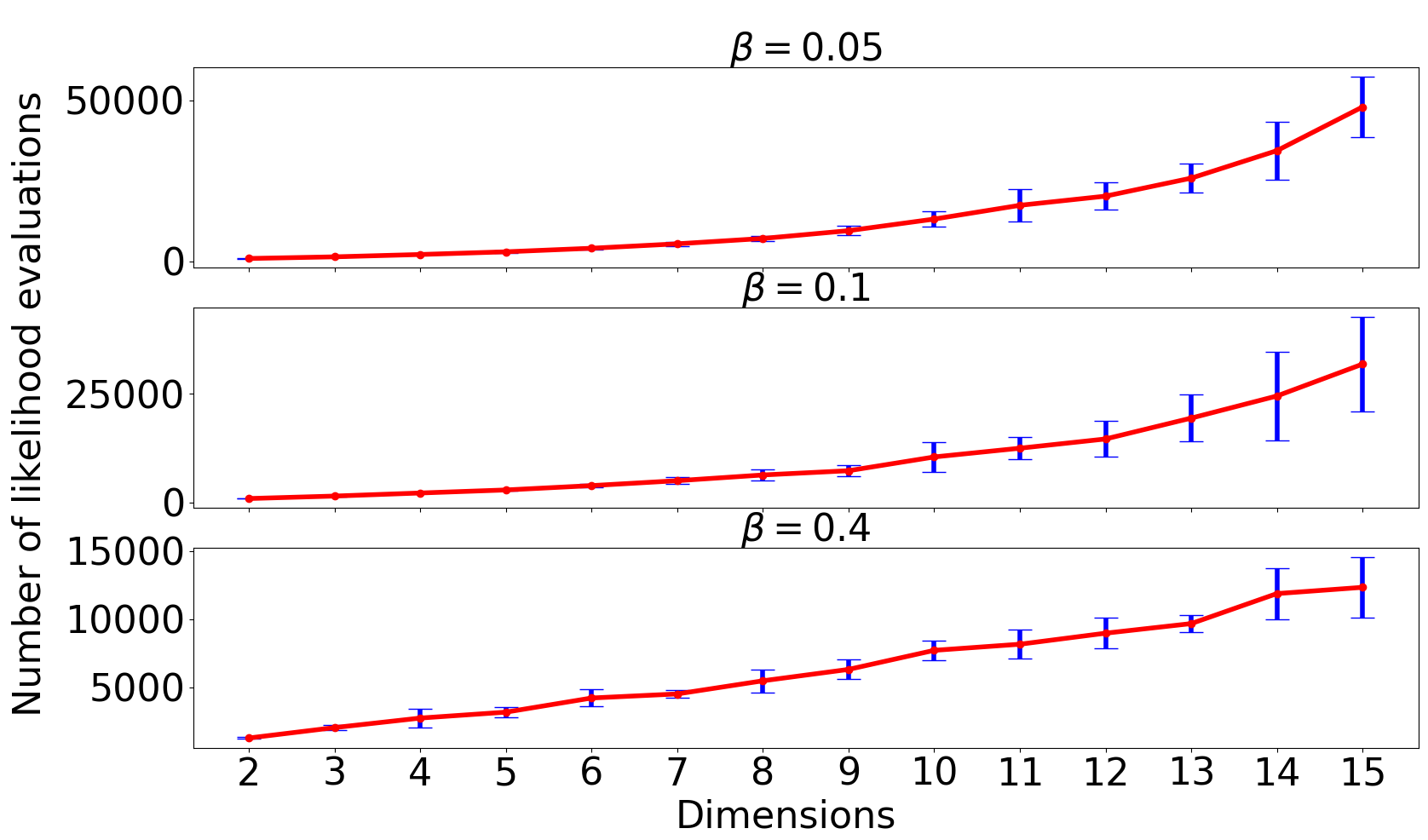}}
\caption{Performance of MultiNest with the PR method applied to the
  case (2) bivariate toy example extended to higher dimensions.  The
  $\mbeta$ values considered are $0.05, 0.1, 0.4$, from top to bottom
  in each subfigure, respectively. The truth for each dimension is set
  to a same value $\mtheta_\ast=1.5$. The RMSE (left-hand column) and
  number of likelihood evaluations (right-hand column) are calculated
  over 20 repeated realisations with same settings as those in
  bivariate example case (2). The red line represents the mean value
  of the repeated realisations, and the blue error bar indicates the
  standard deviation.}
\label{fig:highD}
\end{figure*}

\subsection{Higher-dimensional examples}

In order to investigate the performance of PR in higher
dimensionality, we reconsider case (2) in the bivariate example, but
extend the dimensionality over the range 3 to 15 dimensions.  In
particular, we consider the performance with $\beta = 0.05, 0.1$, and
$0.4$. Each of the experiments is repeated 20 times, and the test
results are evaluated by calculating the mean and standard deviation
of the RMSE over these 20 realisations.

As shown in Figure \ref{fig:highD} (a), with an increase of
dimensionality, the RMSE error-bar undergoes an obvious increase for
both $\mbeta=0.05$ and $0.1$ cases. For the case $\mbeta = 0.4$, the
RMSE increases at lower dimensionality, but then remains at a stable
level for higher dimensionality. Overall, the RMSE performance in
higher dimensions is consistent with that in the bivariate example in
terms of its order of magnitude, which demonstrates that the PR method
is stable and effective for problems with higher dimensionality.

Figure \ref{fig:highD} (b) shows a set of equivalent plots for the
number of likelihood evaluations. This clearly shows that for a
smaller $\mbeta$ value MultiNest makes a larger number of likelihood
evaluations. This is not surprising as a smaller $\mbeta$ corresponds
to a broader \revised{modified} prior space. We note that the number of likelihood
evaluations required for $\mbeta = 0.05$ is almost twice that for
$\mbeta = 0.1$.

\begin{figure}[ht]
\centering
\includegraphics[width = 0.94\linewidth]{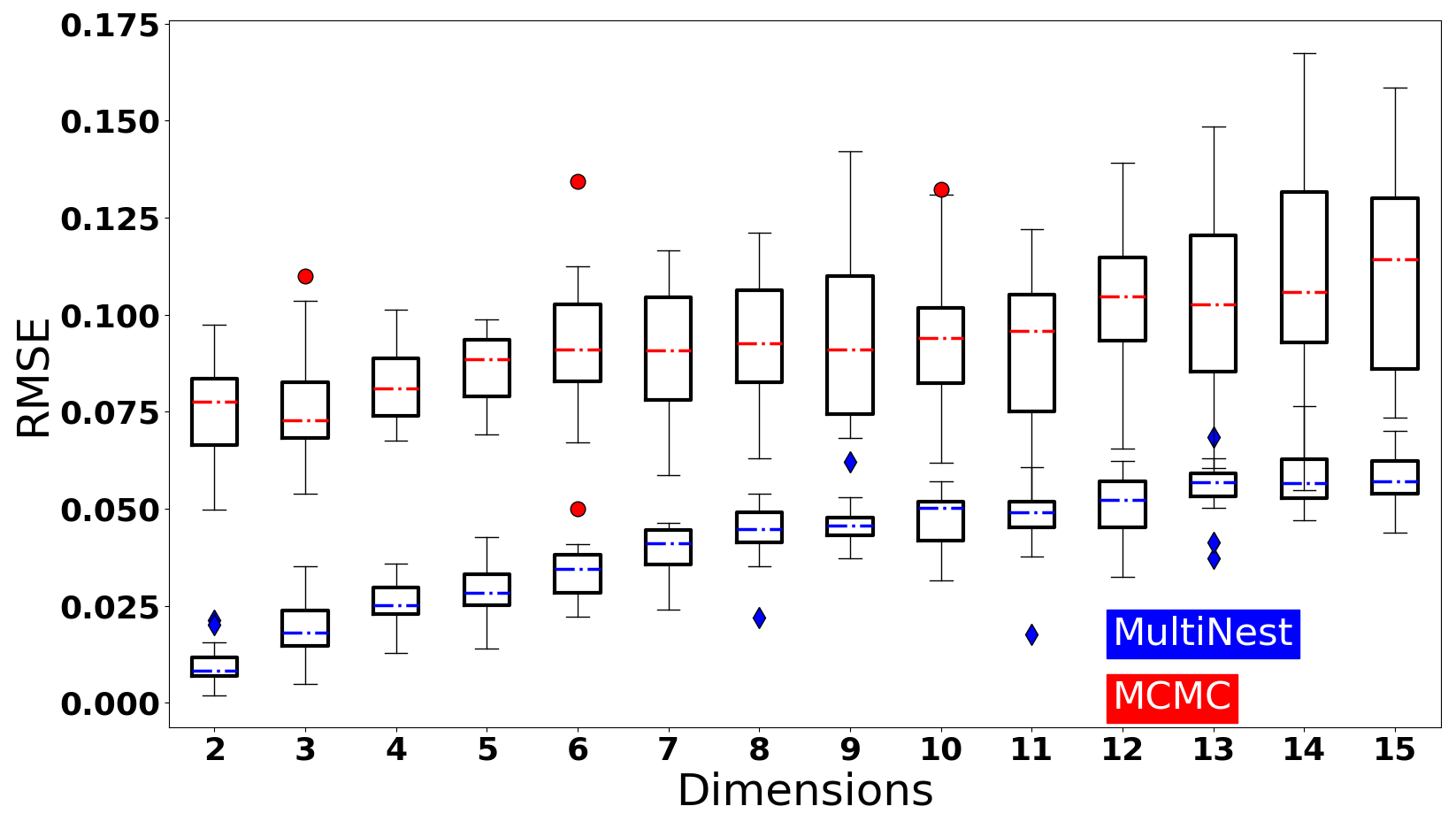}
\caption{RMSE boxplot for high dimensionality comparison between
  MultiNest with the PR method (100 live samples, $\mbeta = 0.05$) and
  MCMC for case (2) $\mtheta_\ast=1.5$. The boxes range from the 25th
  to 75th quantiles. MultiNest results are in blue, and MCMC in
  red. The blue and red dashed lines within the box are the median
  RMSE over 20 realisations for each method. The blue diamond and
  red solid circles represent outliers among the 20 realizations. For each
  dimension, the two methods are computed with a comparable number of
  likelihood evaluations.}
\label{fig:highDComp}
\end{figure}

Figure \ref{fig:highDComp} shows the RMSE comparison between MultiNest
with PR $(\mbeta=0.05)$ and MCMC methods for the same higher dimensional
examples. Note that the RMSE is computed using a comparable
number of likelihood evaluations for the two methods for each
dimensionality. As can be observed from the figure, MCMC remains
stable and accurate (albeit with a slight increase in RMSE with
dimension), but has a higher RMSE than MultiNest with PR across the
dimensionalities considered. By contrast, for MultiNest with PR, the
RMSE increases more noticably with the number of dimensions, as might
be expected from a NS algorithm that is based on a form
rejection sampling.

\begin{table*}[ht]
\centering
\caption{\revised{The performance of MultiNest with and without PR
    method (for various values of $\mbeta$, and both 100 live samples)
    for all three cases of the non-Gaussian bivariate example. The top
    half of the table is a comparison of RMSE, and the second half is
    the number of likelihood evaluations ($N_{\rm like}$) per
    individual algorithm run. For $\mbeta = 0.05$ (highlighted in
    bold) MultiNest achieves relatively better and consistent RMSE and
    $N_{\rm like}$ performance across the three cases.}}
\begin{tabular}{c|cccccc}
\hline
RMSE & MN ($\mbeta=1$)  & $\mbeta=0.4$ & $\mbeta=0.2$ &
$\mbeta=0.1$ & $\mbeta=0.05$ & $\mbeta=10^{-5}$ \\
\hline
Case (1) &  0.0091   &  0.0085 & 0.0065 &  0.0067 & \textbf{0.0066} & 0.0055 \\
Case (2) &  0.2042   &  0.0655 & 0.0186 &  0.0136 & \textbf{0.0135} &  0.0125 \\
Case (3) &  0.1403   &  0.2057 &  0.0705 &  0.0207 & \textbf{0.0196} & 0.0191 \\
\hline
$N_{\rm like}$  \\
\hline
Case (1) &  949   &  926 &  987  &   1049 &  \textbf{1117} & 1313 \\
Case (2) & 2017   &  1356 &  1143  &   1068 &  \textbf{1047} & 1151  \\
Case (3) &  2921  &  1337 &  1067  &   889 &  \textbf{858} & 996 \\
\hline
\end{tabular}
\label{tab:exp2NonGaussian}
\end{table*}

\revised{
\subsection{Non-Gaussian bivariate example}
As our final numerical example, we consider a non-Gaussian bivariate
likelihood function. In particular, we adapt the Gaussian bivariate
likelihood considered in Section~\ref{sec:bivariate} by replacing the
product of Gaussian distributions in each dimension by a product of
Laplace distributions $\mbox{Laplace}(\mu, b)$, so that in each
dimension the Gaussian form (\ref{eqn:likelihood}) is re-written as:
\begin{align}
\mL(\mtheta) = \prod_{n=1}^{N} 
\left\{ \frac{1}{2b} \exp\left(-\frac{|\mtheta - x_n|}{b}\right)
\right\},
\end{align}
where $x_n$ is the $n$th measurement (although  $N=1$ in this example),
which acts as the location parameter similar to in a Gaussian
distribution, and $b$ is the scale parameter in the Laplace
distribution analogous to $\sigma_\xi$ in (\ref{eqn:likelihood}). We
choose the Laplace distribution as our non-Gaussian test example
since: (1) it is valid for both positive and negative values of the
parameter $\theta$, unlike Beta/Gamma distributions; and (2) a Laplace
distribution with a small $b$-value has a similar tail to that of a
Gaussian (i.e. it is not heavy-tailed), which facilitates easier
comparison.}

\revised{The prior distribution is identical to that used
  Section~\ref{sec:bivariate}, i.e. the same Gaussian distribution.
  Indeed, all of the other experimental settings are kept the same as
  those in Section~\ref{sec:bivariate}, and we again consider
  MultiNest with and without PR method in all three cases.}

\revised{The results of the analysis are given in
  Table~\ref{tab:exp2NonGaussian} for runs with $N_{\rm
    live}=100$. Comparing the $N_{\rm live}=100$ results with the
  corresponding ones given in Table~\ref{tab:exp2} for the Gaussian
  bivariate example, ones sees that the trends for both RMSE and
  $N_{\rm like}$ are similar to those in the Gaussian bivariate cases,
  but are in general higher for the Laplace distribution.  This is
  because the peak of the Laplace distribution is sharper than that of
  a Gaussian. Again reasonable results are obtained for $\mbeta =
  0.05$.}

\begin{figure}[ht]
\centering
\includegraphics[width = 0.95\linewidth]{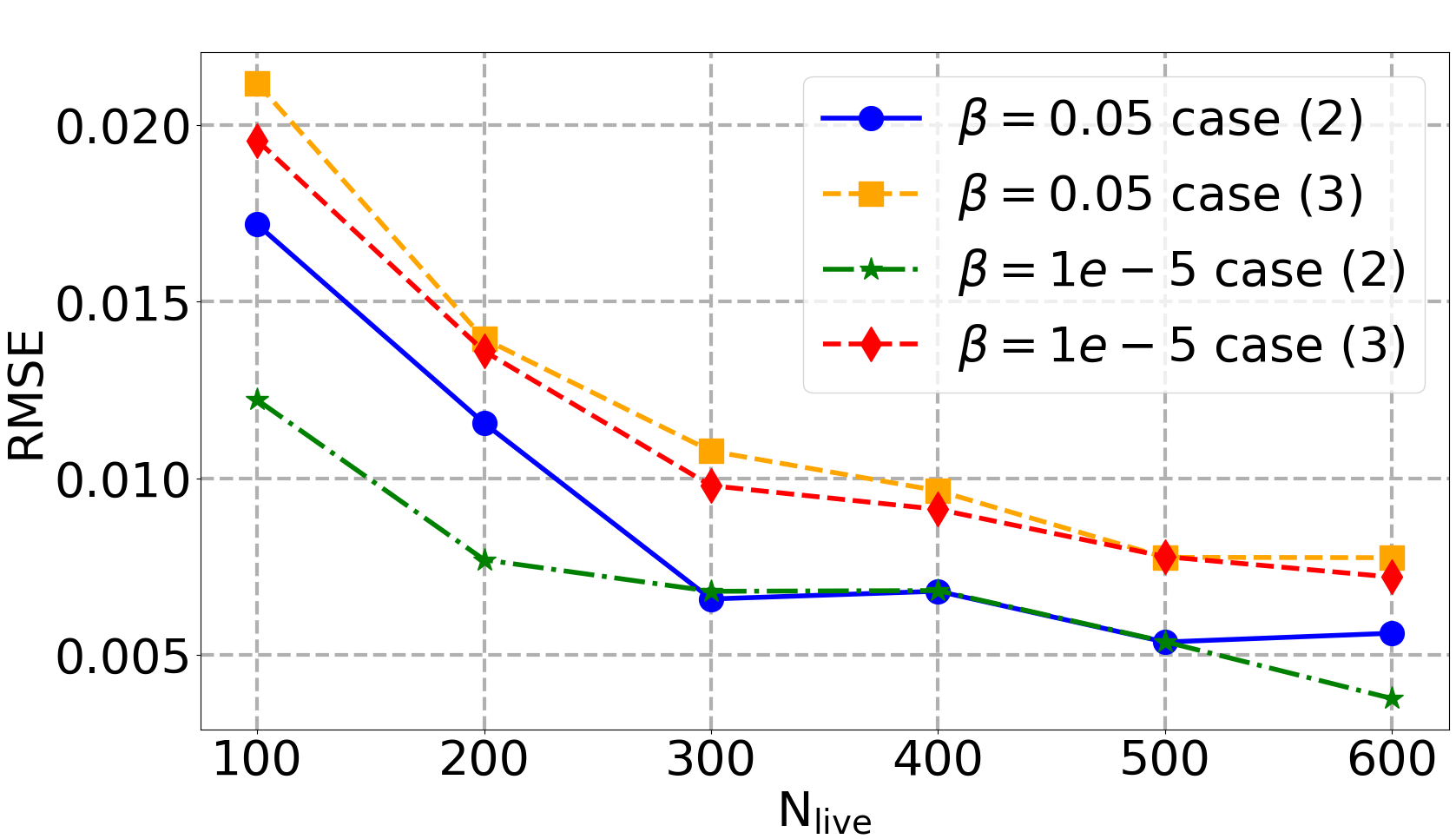}
\caption{\revised{RMSE performance of MultiNest in the non-Gaussian
    bivariate example with different $N_{\rm live}$ and $\mbeta$
    values for case (2) and (3).}}
\label{fig:Exp2nonG_RMSE}
\end{figure}

\revised{Figure \ref{fig:Exp2nonG_RMSE} shows the RMSE resulting from
different $N_{\rm live}$ values for $\mbeta = 0.05$ and $10^{-5}$, respectively.
Comparing these RMSE values with those given in Table~\ref{tab:exp2},
which were obtained for the Gaussian bivariate example with $N_{\rm live}=100$,
one sees that higher $N_{\rm live}$ values are required for the
Laplace distribution to achieve similar levels of accuracy.}

\section{Conclusions} \label{Sec:Con}

This paper addresses the unrepresentative prior problem in Bayesian
inference problems using NS, by introducing the posterior
repartitioning method.

The key advantages of the method are that: (i) it is general in nature
and can be applied to any such inference problem; (ii) it is simple to
implement; and (iii) the posterior distribution is unaltered and hence
so too are the inferences. The method is demonstrated in univariate
and bivariate numerical examples on Gaussian posteriors, and its
performance is further validated and compared with MCMC sampling
methods in examples up to 15 dimensions. \revised{The method is also
  tested on a non-Gaussian bivariate example. In all cases, we
  demonstrate that NS algorithms, assisted by the PR method, can
  achieve accurate posterior estimation and evidence approximation in
  problems with an unrepresentative prior.}

\revised{The proposed scheme does, however, have some limitations: (i)
  if the prior and likelihood are extremely widely separated, the
  sampling can still be inefficient and slow, because of the large
  augmented search space for very small $\mbeta$; (ii) the approach
  cannot be readily applied to problems with discrete parameters; and
  (iii) the normalisation of the modified prior will in general not be
  possible analytically, but require numerical integration.}

\begin{acknowledgement}
The authors thank Dr Detlef Hohl for reviewing the draft and
providing helpful comments, Jaap Leguijt for numerous
useful discussions in the early stages of this work, and Dr Will
Handley for interesting discussions of nested sampling.
\end{acknowledgement}

\bibliographystyle{spbasic}      
\bibliography{bibFile}   

\begin{thebibliography}{19}
\providecommand{\natexlab}[1]{#1}
\providecommand{\url}[1]{{#1}}
\providecommand{\urlprefix}{URL }
\expandafter\ifx\csname urlstyle\endcsname\relax
  \providecommand{\doi}[1]{DOI~\discretionary{}{}{}#1}\else
  \providecommand{\doi}{DOI~\discretionary{}{}{}\begingroup
  \urlstyle{rm}\Url}\fi
\providecommand{\eprint}[2][]{\url{#2}}

\bibitem[{Allison and Dunkley(2013)}]{allison2013comparison}
Allison R, Dunkley J (2013) {Comparison of sampling techniques for Bayesian
  parameter estimation}. Monthly Notices of the Royal Astronomical Society
  437(4):3918--3928

\bibitem[{Bishop(2006)}]{bishop2006}
Bishop C (2006) Pattern recognition and machine learning. Springer

\bibitem[{Chopin and Robert(2010)}]{chopin2010properties}
Chopin N, Robert C (2010) Properties of nested sampling. Biometrika
  97(3):741--755

\bibitem[{Endres and Schindelin(2003)}]{endres2003}
Endres D, Schindelin J (2003) A new metric for probability distributions. IEEE
  Transactions on Information theory 49(7):1858--1860

\bibitem[{Feroz and Hobson(2008)}]{feroz2008multimodalns}
Feroz F, Hobson M (2008) {Multimodal nested sampling: an efficient and robust
  alternative to Markov Chain Monte Carlo methods for astronomical data
  analyses}. Monthly Notices of the Royal Astronomical Society 384(2):449--463

\bibitem[{Feroz et~al(2009)Feroz, Hobson, and Bridges}]{feroz2009multinest}
Feroz F, Hobson M, Bridges M (2009) {MultiNest: an efficient and robust
  Bayesian inference tool for cosmology and particle physics}. Monthly Notices
  of the Royal Astronomical Society 398(4):1601--1614

\bibitem[{Feroz et~al(2013)Feroz, Hobson, Cameron, and
  Pettitt}]{feroz2013importance}
Feroz F, Hobson M, Cameron E, Pettitt A (2013) {Importance nested sampling and
  the MultiNest algorithm}. arXiv preprint arXiv:13062144

\bibitem[{Gelman(2008)}]{gelman2008}
Gelman A (2008) {Objections to Bayesian statistics}. Bayesian Analysis
  3(3):445--449

\bibitem[{Gronau et~al(2017)Gronau, Sarafoglou, Matzke, Ly, Boehm, Marsman,
  Leslie, Forster, Wagenmakers, and Steingroever}]{gronau2017tutorial}
Gronau QF, Sarafoglou A, Matzke D, Ly A, Boehm U, Marsman M, Leslie DS, Forster
  JJ, Wagenmakers EJ, Steingroever H (2017) A tutorial on bridge sampling.
  Journal of mathematical psychology 81:80--97

\bibitem[{Handley et~al(2015)Handley, Hobson, and
  Lasenby}]{handley2015polychord}
Handley W, Hobson M, Lasenby A (2015) {POLYCHORD: next-generation nested
  sampling}. Monthly Notices of the Royal Astronomical Society
  453(4):4384--4398

\bibitem[{Jahn et~al(2018)Jahn, Beaujean, and Straub}]{Jahn2018pypmc}
Jahn S, Beaujean F, Straub D (2018) pypmc. \doi{10.5281/zenodo.1158068},
  \urlprefix\url{https://doi.org/10.5281/zenodo.1158068}

\bibitem[{MacKay(2003)}]{mackay2003}
MacKay D (2003) Information theory, inference and learning algorithms.
  Cambridge university press

\bibitem[{Martino et~al(2018)Martino, Elvira, and
  Camps-Valls}]{martino2017group}
Martino L, Elvira V, Camps-Valls G (2018) {Group Importance Sampling for
  particle filtering and MCMC}. Digital Signal Processing 82:133--151

\bibitem[{Neal(2001)}]{neal2001annealed}
Neal RM (2001) Annealed importance sampling. Statistics and computing
  11(2):125--139

\bibitem[{Salvatier et~al(2016)Salvatier, Wiecki, and
  Fonnesbeck}]{salvatier2016}
Salvatier J, Wiecki T, Fonnesbeck C (2016) {Probabilistic programming in Python
  using PyMC3}. PeerJ Computer Science 2:e55

\bibitem[{Simpson et~al(2017)Simpson, Rue, Riebler, Martins, S{\o}rbye
  et~al}]{simpson2017penalising}
Simpson D, Rue H, Riebler A, Martins TG, S{\o}rbye SH, et~al (2017) Penalising
  model component complexity: A principled, practical approach to constructing
  priors. Statistical Science 32(1):1--28

\bibitem[{Skilling(2006)}]{skilling2006nested}
Skilling J (2006) {Nested Sampling for General Bayesian Computation}. Bayesian
  Analysis 1(4):833--860

\bibitem[{S{\"u}li and Mayers(2003)}]{suli2003}
S{\"u}li E, Mayers D (2003) An introduction to numerical analysis. Cambridge
  university press

\bibitem[{Tokdar and Kass(2010)}]{tokdar2010importance}
Tokdar ST, Kass RE (2010) Importance sampling: a review. Wiley
  Interdisciplinary Reviews: Computational Statistics 2(1):54--60

\end{thebibliography}

\end{document}